\documentclass[aps,twocolumn,pra,showpacs,amsmath,amssymb,longbibliography,showkeys,10pt]{revtex4-2}
\usepackage{graphicx}
\usepackage{epstopdf}
\usepackage[normalem]{ulem}
\usepackage{braket}
\usepackage{hyperref}
\usepackage[dvipsnames]{xcolor}
\usepackage{ulem}
\allowdisplaybreaks

\begin{document}

\title{Statistical properties of quantum jumps between macroscopic states of light: reading an operational coherence record}

\author{Th. K. Mavrogordatos}
\email{themis.mavrogordatos@fysik.su.se}
 \affiliation{Department of Physics, AlbaNova University Center, SE 106 91, Stockholm, Sweden}

\date{\today}

\begin{abstract}
We propose an experimental apparatus to reveal the quantum coherence manifested in downward quantum jumps of amplitude bistability. The underlying coherent superposition of macroscopic quantum states is translated into the statistical properties of the integrated charge deposited in the detector circuit of a mode-matched heterodyne/homodyne detection scheme. At first, the dynamical evolution of a signal transmitted from an auxiliary cavity is employed to pinpoint a macroscopic switching event in a bistable main cavity subject to direct photodetection. Once the decision is made on the occurrence of a downward switch, the main cavity mode is let to freely decay to the vacuum, monitored to the production of an integrated charge. In the long-time limit, the charge distribution over an identical collection of pure states generated during the jumps converges to the $Q$ function (heterodyne detection) or marginals of the Wigner function (homodyne detection) dictated by the phase of the local oscillator. When fluctuations over the ensemble step in, we connect the statistical properties of several switching events and the ensuing production of current records, to the cavity field correlations associated with the breakdown of photon blockade.  
\end{abstract}


\maketitle

\section{Introduction}

The exquisite control recently acquired over cavity and circuit QED architectures~\cite{Carmichael1993QTIV, CarmichaelBook2, Dykman2012, Blais2021, RiceBook,GarcíaRipoll2022} enables the investigation of the nature of quantum jumps~\cite{Bohr1913, Einstein1916, Einstein1917, Schrodinger1952, Plenio1998}, in line with Bohr's insistence that coherence must have a place in the interaction of light with matter~\cite{BKS1924I, BKS1924II, Mabuchi1996, Deleglise2008, CarmichaelBook2}. Further to driven multilevel systems (including atoms, ions, molecules and quantum oscillators among others)---be it natural~\cite{Dehmelt1975, Nagourney1986, Sauter1986, Bergquist1986, Basche1995, Carmichael1997} or artificial~\cite{Peil1999, Vijay2011, Minev2019}---with some ever {\it stable} energy states, new prospects in the study of multi-photon quantum-nonlinear optics are opened by considering stochastic switching between {\it metastable states} of light, where often very large numbers of photons are involved in the display of criticality~\cite{Sett2024}. Such states are dynamically formed and influenced in a fundamental way by their inputs and outputs. The breakdown of photon blockade via a first-order quantum phase transition in the open Jaynes--Cummings (JC) model~\cite{JaynesCummings1963, Paul1963} driven out of resonance~\cite{Carmichael2015, Fink2017}, places amplitude bistability~\cite{Narducci78, Bonifacio1978, Savage1988, Armen2009, Mabuchi2018} at a prominent position as a nonlinear response averaged over quantum fluctuations, whence rendering increasingly relevant the question where can one look for signs of quantum coherence in the switching between macroscopic states. Similarly, are we able to perform a measurement which correlates the stochastic aspect of bistable switching to the conditioned yet deterministic preparation of a quantum state with unequivocal coherence properties?

Motivated by the ``stabilization of classical attractors'' reported in recent circuit QED experiments on quantum amplitude bistability where the light-matter coupling strength is largely in excess of the dissipation rates~\cite{Fink2017, Sett2024}, in conjunction with the remarkable coherence times of superconducting qubits~\cite{Houck2008, Place2021, Blais2021, GarcíaRipoll2022}, in this report we set out to answer these questions in an operational fashion. To this aim, we propose an experimental design deploying complementary methods of making scattering records~\cite{Carmichael1993QTII, Carmichael1999, CarmichaelBook2, Wiseman2012}. Using quantum trajectory theory, we will explore the statistical properties of macroscopic quantum jumps in terms of the dynamical macroscopic superposition states~\cite{Schrodinger1935, Dodonov1974, Gerry1997, HarocheBook, Nielsen2006, Ourjoumtsev2006, Ourjoumtsev2007, Sychev2017, Hacker2019,Pan2023} they condition, and their tomography produced by homodyne and heterodyne detection. In conjunction with ensemble-averaged quantities, we will also look at individual realizations uncovering contextual wave/particle correlations of the light radiated from a cavity initialized in the above conditioned states. 

The two-step protocol is laid out in Sec.~\ref{sec:protocol}, probing the fluctuations of the quantum radiation field emanating from a paradigmatic nonlinear oscillator---the JC source. Unraveling the underlying master equation by means of direct photodetection substantiates the first step of the experiment and is briefly discussed in Sec.~\ref{sec:ME}; this first step draws heavily on the analysis of Ref.~\cite{JumpsBist}. However, to substantiate a readout scheme, an auxiliary system is warranted for the translation of the quantum coherence associated with a downward switching event into a trigger initiating the next step and activate a homodyne/heterodyne detection circuitry. Consequently, the second step relies on complementary methods of record making to provide an operational tomography of the anterior macroscopic quantum jumps, a process detailed in Sec.~\ref{sec:compmethods}. The statistical properties of the collected ensemble reflecting systematic errors and inherent quantum fluctuations associated with the jump process are imprinted on the interference fringe visibility pattern, as we briefly discuss in Sec.~\ref{sec:statdev}. Moreover, the commonly encountered experimental constraint of limited detection efficiency motivates the consideration of conditioned homodyne detection Sec.~\ref{sec:wp}, where the measured signal reveals the correlation of the photon number with a set quadrature amplitude of the light field. The source is, as before, a freely dissipated mode supported by a lossy cavity, initialized in the conditioned state of a quantum jump. In this case, however, the contextual reading of the output determines the timescale over which signatures of coherence can be sustained by producing large wave/particle correlations of varying sign. Short concluding remarks summarize the findings to close the paper out.  

\section{A quantum non-demolition protocol for the source-field fluctuations}
\label{sec:protocol}

\begin{figure*}
\includegraphics[width=0.83\textwidth]{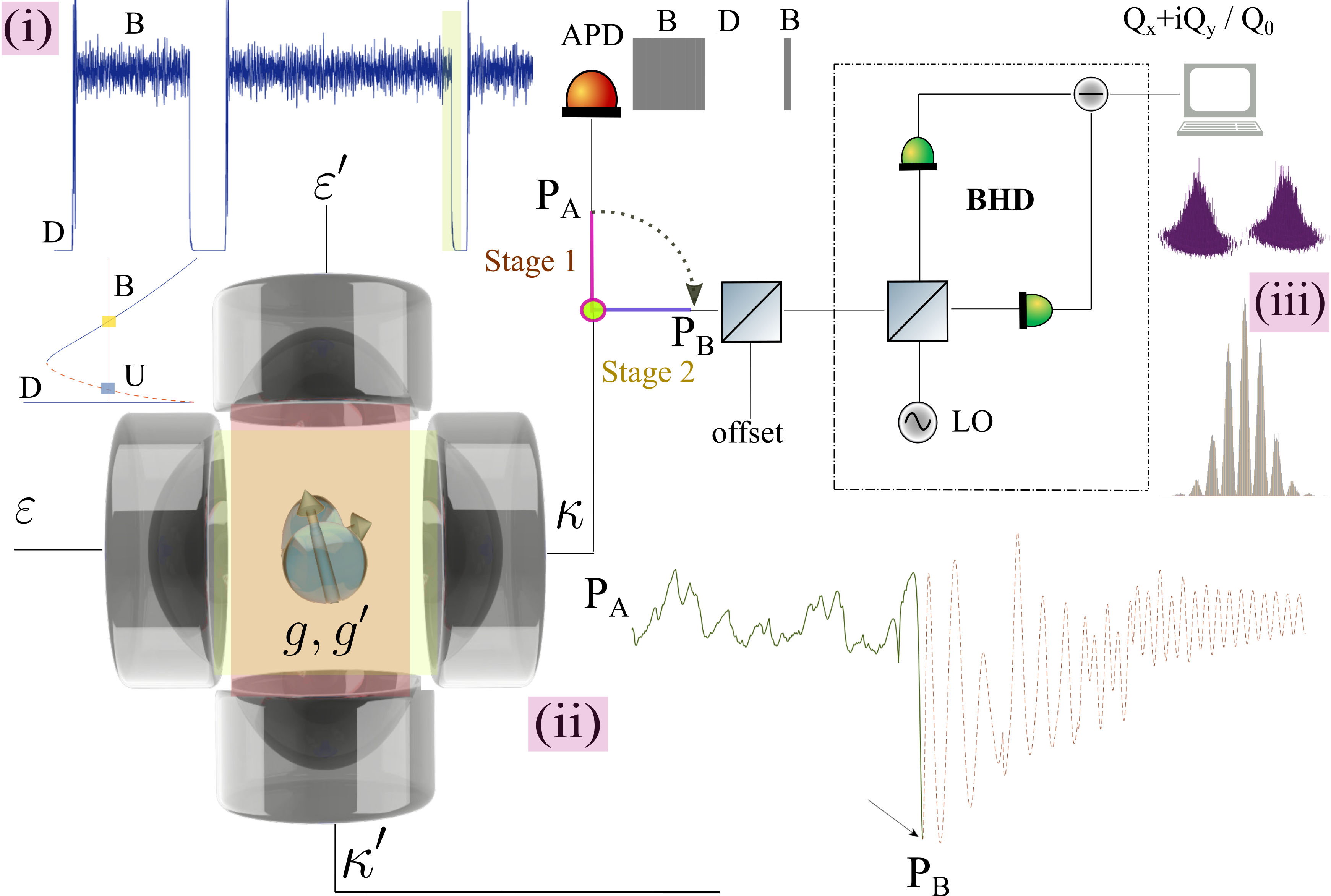}
\caption{{\it Schematic illustration of the proposed experimental setup.} The main cavity mode is coherently driven with drive amplitude $\varepsilon$ and strongly coupled to a two-state atom with strength $g$. Photon loss at rate $2\kappa$ establishes the only channel for leaking information to the environment, where an avalanche photodiode (APD) detector (or its microwave analogue) is placed. The auxiliary cavity mode is driven with amplitude $\varepsilon^{\prime}$ and is weakly coupled to the main cavity mode with strength $g^{\prime}$. The rapid decay from the auxiliary cavity determines the dominant rate $\kappa^{\prime}$ allowing for adiabatic elimination of its resonating mode field. The main cavity-atom system is operated under conditions of high-amplitude bistability with a typical (schematic) realization of the conditioned output flux rate pictured in inset (i) for $800$ cavity lifetimes. For the control switch at the position ${\rm P_A}$ {\bf (Stage 1)}, the record obtained under direct photodetection consists of the alternation between time periods of closely spaced photoelectron ``clicks'', corresponding to the bright (B) state, and periods of several cavity lifetimes with no ``clicks'', corresponding to the dim (D) state. The latter two states, together with the unstable (U) state are also identified from the semiclassical hysteresis curve illustrated below the main trajectory of (i). Inset (ii) depicts the signal transmitted from the auxiliary cavity during a downward quantum jump, focused about the green-shaded area of the trajectory (i). It shows the dip dictating when the control switch is to be moved to position ${\rm P_B}$, indicated by the arrow, actuating balanced and mode-matched homodyne/heterodyne detection {\bf (Stage 2)}. In the balanced detection scheme (BHD) two fields are superposed past the second beam splitter, the cavity output field and the local-oscillator field, with a $\pi$ phase difference between the superpositions. The integrated charge deposited in the detector circuit as a result of their detection~\cite{CarmichaelBook2} is $Q=Q_x+iQ_y$ for heterodyne and $Q_{\theta}$ for homodyne detection. Inset (iii) depicts a steady-state distribution $P(Q^{*},Q)$ (top) and $P(Q_{\theta})$ (bottom) over an ensemble average of realizations with a coherent-state superposition as a common start for the cavity state.  The coherent offset (applied at the first beam splitter) is set to zero for heterodyne and to $-(\alpha_1+\alpha_2)/2$ for homodyne detection, where $\alpha_1$ and $\alpha_2$ are the conditioned coherent state amplitudes attributed to the bright and unstable states, respectively.}
\label{fig:FIG1}
\end{figure*}

A schematic of the suggested experimental configuration is depicted in Fig.~\ref{fig:FIG1}, featuring two principal stages. The two stages are temporally separated by a control switch that moves from position ${\rm P_A}$ to the position ${\rm P_B}$ terminating \underline{Stage 1}, during which the macroscopic quantum jumps of interest occur, and initiating \underline{Stage 2}, during which the operational consequences of these monitored jumps play out. Both stages are amenable to the theory of quantum trajectories~\cite{Belavkin1990, Barchielli1991, Alsing1991, Dalibard1992, CarmichaelBook2} applied here to treat two explicitly open quantum systems: a JC complex coupled to an auxiliary cavity, both subject to photon loss leaking information to the environment, and, subsequently, a single cavity mode decaying to the vacuum. The said theory provides the connection between measurement records and the conditioned wavefunction evolution. Throughout \underline{Stage 1}, we are dealing with the source, a main cavity mode strongly coupled to a two-level `atom',  weakly coupled to an auxiliary cavity field, acting as a meter system. We will shortly find that the bad-cavity limit of the auxiliary cavity substantiates a quantum non-demolition (QND) measurement for the main one~\cite{Peil1999, Guerlin2007,Sun2014, Minev2019}. We are thus employing the open JC model with the main field mode and the `atom' on resonance, with frequency $\omega_0$. The main cavity mode is coherently driven with a field of frequency $\omega_d$, in resonance with the auxiliary cavity mode. 

In a frame rotating with the drive, the Lindblad master equation (ME) for the density matrix $\varrho$ of the composite system under study may be written in the form~\cite{Carmichael1993QTI}
\begin{equation}\label{eq:MEfull}
\frac{d\varrho}{dt}=\frac{1}{i\hbar}[H, \varrho] + \mathcal{L}[\sqrt{2\kappa}\,a]\varrho + \mathcal{L}[\sqrt{2\kappa^{\prime}}\,a^{\prime}]\varrho,
\end{equation}
with $\mathcal{L}[\xi] \cdot \equiv \xi \cdot \xi^{\dagger}-\tfrac{1}{2}(\xi^{\dagger}\xi \cdot +\cdot \xi^{\dagger}\xi)$ and
\begin{equation}
\begin{aligned}
H&=-\hbar \Delta \omega(a^{\dagger}a + \sigma_{+}\sigma_{-})+i\hbar g (a^{\dagger}\sigma_{-}-a\sigma_{+}) \\
&+ \hbar (\varepsilon a^{\dagger} + \varepsilon^{*}a) + i\hbar g^{\prime}(a^{\prime\dagger}a-a^{\dagger}a^{\prime}) + i \hbar (\varepsilon^{\prime} a^{\prime\dagger} - \varepsilon^{\prime *}a^{\prime}).
 \end{aligned}
\end{equation}
In the above RWA (rotating wave approximation) Hamiltonian, $a^{\dagger}$ and $a$ ($a^{\prime\dagger}$ and $a^{\prime}$) are the creation and annihilation operators for the main (auxiliary) cavity field, $\sigma_{+}$ and $\sigma_{-}$ are the raising and lowering operators for the `atom', and $g$ is the dipole coupling strength. The quadrature phase operators of the main cavity field are defined as $\mathcal{A}_{\theta}\equiv\frac{1}{2} (a e^{-i\theta} + a^{\dagger}e^{i\theta})$ and similarly for the auxiliary cavity ($\mathcal{A}^{\prime}_{\theta}$). The main cavity mode is coherently driven by an external field and, at the same time, decays to the environment at an average rate $\kappa \ll g$; this condition defines the strong-coupling limit of the open quantum system in the ``zero system size'' limit~\citep{CarmichaelBook2}. The former is considered as a formal limit~\cite{Alsing1991} where the `atomic' spontaneous emission rate $\gamma$ over the photon loss rate $2\kappa$ tends to zero [$\gamma/(2\kappa) \to 0$], which is at present routinely achieved in circuit QED architectures~\cite{Houck2008, Place2021}. 

The main cavity output is continuously monitored via {\it direct} photoelectron detection~\citep{Carmichael1993QTI}. The drive detuning $\Delta\omega \equiv \omega_d-\omega_0$ along the coherent drive amplitude $\varepsilon$ are selected so as to lead to high-excitation quantum amplitude bistability evinced in individual realizations, such as the sample displayed in inset (i) of Fig.~\ref{fig:FIG1}. Moreover, the main and auxiliary cavity modes are weakly and linearly coupled with strength $g^{\prime} \ll g$ in a standard beam-splitter type interaction~\cite{Gao2018, Zhang2019, Chapman2023, Bembenek2025}. The rate $\kappa^{\prime}$ at which the auxiliary cavity decays satisfies
\begin{equation}\label{eq:conditions}
\kappa^{\prime} \gg \kappa, g^{\prime} \quad \text{and} \quad g^{\prime 2}/(\kappa \kappa^{\prime}) \ll 1.
\end{equation}
The above two conditions permit the adiabatic elimination of the auxiliary cavity field along the factorization of $\varrho=\rho \otimes \rho^{\prime}$~\cite{BadCavity1988, CarmichaelBook2}. The procedure ultimately results in the reduction of the ME~\eqref{eq:MEfull} to the ME of JC bistability for $\rho$, the latter pertaining to the main cavity mode strongly coupled to the two-state atom. 

With these conditions upheld, the main system density matrix $\rho$ obeys the reduced equation reading
\begin{equation}\label{eq:MEJC}
\frac{d\rho}{dt}=\frac{1}{i\hbar}[H_{\rm JC}, \rho] + \mathcal{L}[\sqrt{2\kappa}\,a]\rho,
\end{equation}
with
\begin{equation}
H_{\rm JC}=-\hbar \Delta \omega(a^{\dagger}a + \sigma_{+}\sigma_{-})+i\hbar g (a^{\dagger}\sigma_{-}-a\sigma_{+}) + \hbar (\varepsilon a^{\dagger} + \varepsilon^{*}a).
\end{equation}
Denoting with $|+\rangle$ and $|-\rangle$ the upper and lower states of the two-level `atom' we can compute the main cavity density matrix as $\rho_c=\langle + |\rho| + \rangle  +\langle - |\rho| - \rangle$ and from it the $Q$ function as a phase-space representation~\cite{CarmichaelBook1}:
\begin{equation}
\mathcal{Q}(x+iy)=e^{-(x^2+y^2)}\sum_{n,m}\frac{(x-iy)^n (x+iy)^m}{\sqrt{m!n!}} \langle n | \rho_c |m \rangle,
\end{equation}
where $|n\rangle$ ($n=0,1,2,\ldots$) are Fock states of the cavity field.

At the same time, following adiabatic elimination, the auxiliary cavity field average is given by the expression
\begin{equation}\label{eq:ava}
\langle a^{\prime}(t)\rangle=\frac{g^{\prime}}{\kappa^{\prime}}\langle a(t)\rangle + \frac{\varepsilon^{\prime}}{\kappa^{\prime}},
\end{equation}
with the ensemble average on the RHS reflecting the JC nonlinearity in the main cavity mode. 

The neoclassical field amplitudes (semiclassical amplitudes are denoted with a tilde sign on top) obey the following equations of motion~\citep{CarmichaelBook2}, obtained by taking expectation values in the Heisenberg equations derived from the ME~\eqref{eq:MEJC}:
\begin{subequations}
\begin{align}
\frac{d\tilde{\alpha}}{dt}&=-(\kappa-i\Delta\omega)\tilde{\alpha}-ig\tilde{\beta}-i \varepsilon,\\
\frac{d\tilde{\beta}}{dt}&=i\Delta\omega\,\tilde{\beta} + i g \tilde{\alpha}\tilde{\zeta},\\
\frac{d\tilde{\zeta}}{dt}&=2i\,g\,(\tilde{\alpha}^* \tilde{\beta}-\tilde{\alpha}\tilde{\beta}^*),
\end{align}
\end{subequations}
with $\tilde{\alpha}=\langle a \rangle$, $\tilde{\beta}=\langle \sigma_{-}\rangle$, and $\tilde{\zeta}=2\langle \sigma_{+}\sigma_{-} \rangle -1$. In the steady state, the mean intracavity amplitude solves the following nonlinear equation: 
\begin{equation}\label{eq:neocl}
\begin{aligned}
&\tilde{\alpha}=-i (\varepsilon/\kappa)\\
& \times \left[1-i\left(\Delta\omega/\kappa \mp \mathrm{sgn}(\Delta\omega) \frac{(g/\kappa)^2}{\sqrt{(\Delta\omega/\kappa)^2 + |\tilde{\alpha}|^2/n_{\rm sat}}}\right) \right]^{-1},
\end{aligned}
\end{equation}
where $n_{\rm sat}\equiv [g/(2\kappa)]^2$ is the saturation photon number. In the strong-coupling limit, a large intracavity excitation is required to turn on the JC nonlinearity in view of $n_{\rm sat} \gg 1$~\cite{Carmichael2015}. For a given value of $\varepsilon/\kappa$ in the regime of amplitude bistability, Eq.~\eqref{eq:neocl} is solved by three states corresponding to three complex amplitudes $\tilde{\alpha}_1, \tilde{\alpha}_2$ and $\tilde{\alpha}_3$. Two of them are stable, namely the {\it bright} state (B) with amplitude $|\tilde{\alpha}_1| \gg 1$ and the {\it dim} state (D) with amplitude $|\tilde{\alpha}_3| \sim 0$. The remaining one with amplitude $|\tilde{\alpha}_2| \sim 1$ is unstable (U) to fluctuations (see Sec. 14.2 of~\cite{CarmichaelBook2}). Only the $B$ and $D$ states can be discerned as peaking distributions in the regime of bimodality identified from the numerical solution of the ME~\eqref{eq:MEJC}. The B$\to$D (D$\to$B) transitions in single quantum trajectories are termed {\it downward (upward)} quantum jumps or switching events, while the two participating states are {\it metastable} with a lifetime largely in excess of $\kappa^{-1}$ when $n_{\rm sat}\to \infty$~\cite{Sett2024}. So far the role of the unstable state in organizing the conditioned quantum dynamics remains elusive. Our task in this report is to operationally uncover the crucial role of the U state in mediating a downward jump, during a time period which is typically a small fraction of the cavity lifetime~\cite{JumpsBist}.
 
For the setup pictured in Fig.~\ref{fig:FIG1}, the semiclassical bistability amplitudes dictate the selection of the amplitude for the external coherent drive to the auxiliary cavity, as $\varepsilon^{\prime}=-g^{\prime} \tilde{\alpha}_1/2$. On the basis of Eq.~\eqref{eq:ava}, we deduce that the auxiliary cavity field is in fact the empty-cavity response to the interference of a quantum and a classical field with combined amplitude $\langle a(t) \rangle-\tilde{\alpha}_1/2$. One clarification, however, is warranted here: both of these fields are no longer ensemble-averaged quantities but are explicitly conditioned on a particular downward switch of amplitude bistability. In the following section, we trace the self-consistent mechanism of this conditioning. 

\section{Unraveling the master equation of Jaynes--Cummings bistability}
\label{sec:ME}

We can now focus on the direct photodetection unraveling of the ME~\eqref{eq:MEJC} in order to generate conditioned averages of common observables revealing the coherence associated with quantum jumps, like the average intracavity field $\langle a(t) \rangle$. We first return to the ME~\eqref{eq:MEfull} for the composite system. Consistent with the process of adiabatic elimination~\cite{CarmichaelBook2} we met in Sec.~\ref{sec:protocol}, the cavity field may be assigned a coherent state with $\rho^{\prime}=\overline{|\alpha^{\prime}\rangle \langle \alpha^{\prime}|}$, where the overbar denotes the ensemble average over individual realizations. 

The amplitude $\alpha^{\prime}$ is screened by the ratio $(g^{\prime}/\kappa^{\prime}) \ll 1$, and satisfies the following stochastic equation, balancing the large auxiliary cavity loss with the stochastic drive arising as a superposition of two fields:  
\begin{equation}
\frac{d\alpha^{\prime}}{dt}=-\kappa^{\prime}\alpha^{\prime} + g^{\prime}(\alpha_t - \tilde{\alpha}_1/2),
\end{equation}
where $\alpha_t \equiv \langle a(t) \rangle_{\rm REC}=\langle \psi_{\rm REC}(t)| a |\psi_{\rm REC}(t)\rangle$ represents an individual-trajectory average of the main cavity field, acting as an effective drive amplitude to the auxiliary cavity (the label ${\rm REC}$ identifies the scattering records of photoelectric detection). These averages are produced by unraveling ME~\eqref{eq:MEJC} under direct photodetection~\cite{Glauber1963, Glauber1963Coh, Glauber1963Rep, Carmichael1993QTI}; a typical record is displayed in Fig.~\ref{fig:FIG1} as the alternation of highly bunched sequences of photoelectron ``clicks'' (B) and of long periods of absent APD firings (D). Sample realizations are generated through a quantum Monte Carlo algorithm (see Supplementary Material of~\citep{JumpsBist}), while the numerical solution of the ME~\eqref{eq:MEJC} is accomplished via exact diagonalization in \textsc{MATLAB}'s Quantum Optics Toolbox~\cite{Tan1999}.    

We note that the above procedure is equivalent with solving a Schr\"{o}dinger equation with the non-Hermitian Hamiltonian
\begin{equation}\label{eq:HnonH}
H^{\prime}=i\hbar (\varepsilon_{\rm tot}^{\prime} a^{\prime\dagger}-\varepsilon_{\rm tot}^{*\prime}a^{\prime}) - i\hbar \kappa^{\prime}a^{\prime \dagger}a^{\prime},
\end{equation}
which is in turn tantamount to formulating a ME evolution, one where the probability for a photon ``click'' registered at the detector in the environment is vanishingly small. In the continuous evolution thus generated, the combined stochastic drive amplitude is $\varepsilon_{\rm tot}^{\prime} \equiv \alpha_t - \tilde{\alpha}_1/2$.

Let us briefly mention some key properties of the stochastic field $\alpha_t$, originating from the main cavity, which interferes with the external coherent drive applied to the auxiliary cavity. In a recent treatment of quantum amplitude bistability~\cite{JumpsBist}, it was argued that there is a notable asymmetry between upward and downward jumps; the latter are in principle subject to a coherent-state localization between the bright and an unstable state, dynamically identified in the course of single trajectories under direct photodetection. Continuous measurement thus conditions a bound pair of macroscopic states (bright and unstable) with correlation time $\Delta t_{\rm loc} \ll \kappa^{-1}$. In the absence of interruption from photoelectron ``clicks'', the downward switch is completed within a lower time bound approximated by the expression~\cite{JumpsBist}
\begin{equation}\label{eq:Dtloc}
\kappa \Delta t_{\rm loc} = -\frac{\ln\left(\displaystyle\frac{|\tilde{\alpha}_2|}{|\tilde{\alpha}_1|^2-|\tilde{\alpha}_2|(|\tilde{\alpha}_2|-1)}\right)}{|\tilde{\alpha}_1|^2-|\tilde{\alpha}_2|^2}.
\end{equation} 
When examining individual quantum trajectories, generated in the regime of high-excitation JC bistability where $|\tilde{\alpha}_1-\tilde{\alpha}_2|^2 \gg 1$, the amplitudes of the bright and unstable states are read from the conditioned {\it quasi}probability distribution of the main cavity field, and are denoted by $\alpha_1$, $\alpha_2$ respectively (without a tilde sign on top). Focusing on the region of a downward switch, the condition $\kappa^{\prime} \gg \kappa$ is put on a firmer basis in light of Eq.~\eqref{eq:Dtloc}: to resolve a quantum jump in the auxiliary cavity driven by the field of the main one, we require $\kappa^{\prime}/\kappa > |\tilde{\alpha}_1|^2/\ln(|\tilde{\alpha}_1|^2) \gg 1$. Moreover, under the conditions~\eqref{eq:conditions}, the strategy followed to measure the auxiliary cavity observables at the discretion of the experimenter, does not impact on the record making process to unravel ME~\eqref{eq:MEJC}. In this sense, a QND measurement of the main cavity state is accomplished at the level of single trajectories. 

We arrive at the central criterion employed to decide whether a downward jump has occurred: subtracting $g \tilde{\alpha}_1/2$ from $\alpha_t$ via coherently driving the auxiliary cavity, ensures that the total drive will cross zero at some point during the localization, precipitating a sharp dip in the scaled cavity transmission or output flux. The latter two quantities are respectively defined as
\begin{equation}\label{eq:TF}
T^{\prime}=(\kappa^{\prime}/g^{\prime})^2 \left|\langle a^{\prime}(t)\rangle \right|^2, \quad F^{\prime}=(\kappa^{\prime}/g^{\prime})^2\langle a^{\prime\dagger}a^{\prime}(t)\rangle.
\end{equation}
In the bad-cavity limit, they both coincide to $F^{\prime}=T^{\prime}=(\kappa^{\prime}/g^{\prime})^2 |\alpha^{\prime}(t)|^2$. Locating the sharp dip at the time $t_{\rm dip}$ in transmission in the course of a single realization concludes \underline{Stage 1} of the experiment.

\begin{figure*}
\includegraphics[width=\textwidth]{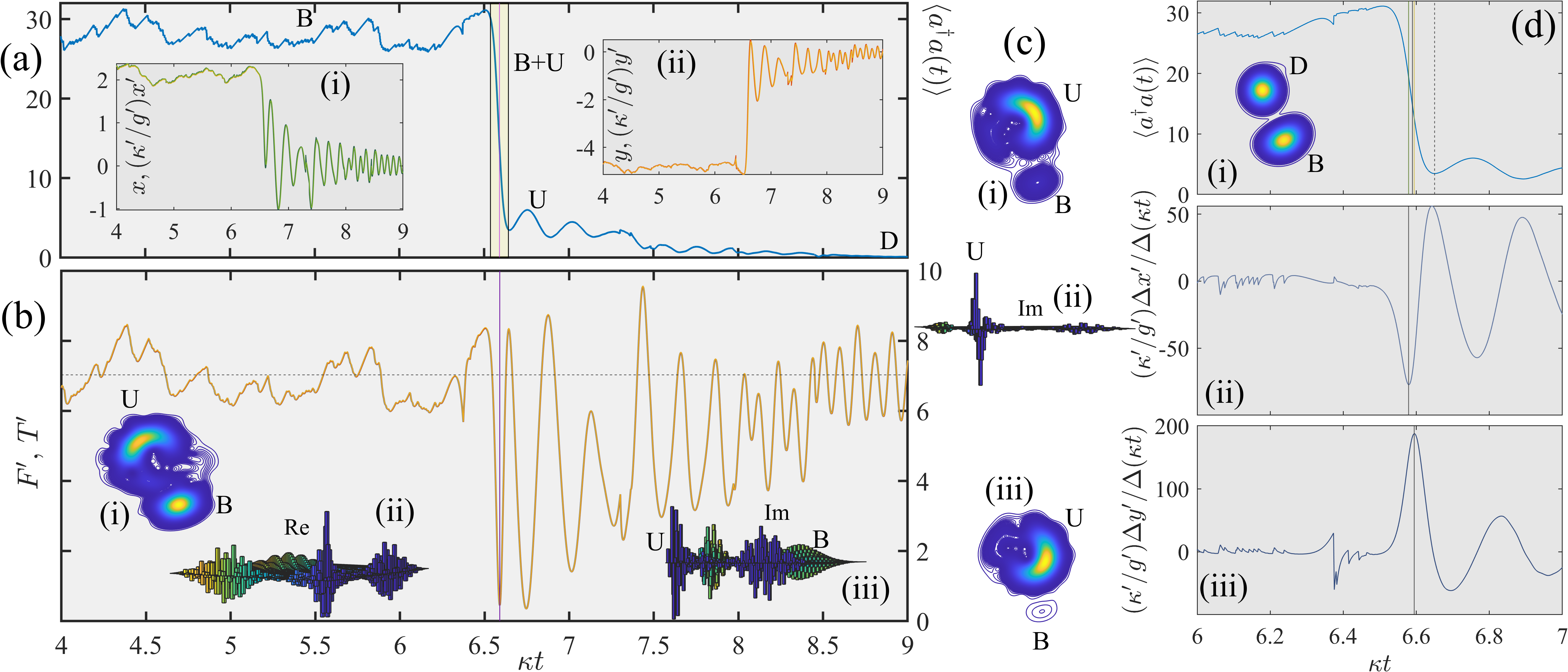}
\caption{{\it Individual realizations, jump detection and initial state preparation.} {\bf (a)} Sample trajectory interval of the main cavity photon number $\langle a^{\dagger}a(t)\rangle=\langle a^{\dagger}a(t)\rangle_{\rm REC}=\langle \psi_{\rm REC}(t)| a^{\dagger}a |\psi_{\rm REC}(t)\rangle$ against five cavity lifetimes centered about a downward switch from the bright (B) to the dim (D) state. The trajectory includes a short interval shown by the yellow shaded area of duration $\kappa\Delta t_{\rm loc}\approx 0.106$ during which coherent localization occurs from the bright state to an unstable state (U) while the two states coexist with varying weights. Inset (i) depicts the quadrature phase amplitude $x=\langle \mathcal{A}_0(t) \rangle_{\rm REC}$ of the main cavity field after a direct detection unraveling of ME~\eqref{eq:MEJC}, superimposed onto the scaled amplitude $(\kappa^{\prime}/g^{\prime})x^{\prime}$ of the auxiliary cavity produced by the solution of a Schr\"{o}dinger equation with Hamiltonian~\eqref{eq:HnonH} and external drive $\varepsilon^{\prime}=0$. Similarly in inset (ii) for $y=\langle \mathcal{A}_{\pi/2}(t) \rangle_{\rm REC}$ and $(\kappa^{\prime}/g^{\prime})y^{\prime}$. {\bf (b)} Scaled outgoing photon flux and transmission from the auxiliary cavity, as defined in Eq.~\eqref{eq:TF}, against the same dimensionless time interval focused upon in (a). The external field amplitude is set to $-\tilde{\alpha}_1/2$, where $\tilde{\alpha}_1\approx 2.082 -4.875i$ solves Eq.~\eqref{eq:neocl}. The purple line in frames (a, b) indicates the position of the dip in flux and transmission, occurring at $\kappa t_{\rm dip}=6.590$ [at the center of the yellow-shaded area in (a)]. The horizontal line marks the position of $|\tilde{\alpha}_1|^2/4 \approx 7.025$. Inset (i) shows an unlabeled contour plot of the conditioned {\it quasi}probability distribution of the main cavity field $\mathcal{Q}(x+iy; t_{\rm dip})$. The peak locations, corresponding to the conditioned coherent state amplitudes are read as $\alpha_1 \approx 1.95-5.45i$ and $\alpha_2 \approx -1.40+0.85i$. Insets (ii), (iii) depict barplots of the real and imaginary parts, respectively, of the conditioned cavity density matrix $[\rho_{c}(t_{\rm dip})]_{mn}$. {\bf (c)} (i) Conditioned {\it quasi}probability distribution $\mathcal{Q}(x+iy; t_{1})$, at $\kappa t_1=6.650$, a time indicated by the dashed line in frame (d, i); (ii) Imaginary part of $[\rho_{c}(t_{1})]_{mn}$, and (iii) $\mathcal{Q}(x+iy; t_{2})$ at $\kappa t_2=6.695$. {\bf (d)} Closer focus on the coherent localization within a cavity lifetime. Frame (i) depicts a close-up on the trajectory of (a). The inset on the left depicts the $Q$ function of the steady-state cavity field $\mathcal{Q}_{\rm ss}(x+iy)$ with photon average $\langle a^{\dagger}a \rangle_{\rm ss}\approx 14.65$ [ensemble average after the solution of ME~\eqref{eq:MEJC}]. In $\mathcal{Q}_{\rm ss}(x+iy)$, the D state is practically the vacuum while the B state peaks at $\approx 2.05-4.85i$. Frames (ii) and (iii) depict the scaled time derivative of the $x$ and $y$-quadratures of the auxiliary cavity field. The two vertical lines mark the position of the minimum and maximum, respectively. These two times, along with $\kappa t_{\rm dip}=6.590$, are all indicated in a cluster of closely-spaced lines in frame (i). The operating parameters read: $g/\kappa=60$, $\varepsilon/\kappa=13.5i$, $\Delta\omega/\kappa=-8$, and $\kappa^{\prime}/\kappa=\kappa^{\prime}/g^{\prime}=100$.}
\label{fig:FIG2}
\end{figure*}

Figure~\ref{fig:FIG2} is devoted to key aspects of quantum coherence associated with a downward switching event of amplitude bistability. Operating conditions produce a state coexistence in the ensemble-averaged distribution of the cavity field, typical of the first-order quantum phase transition via which photon blockade breaks down~\cite{Carmichael2015}. Frame (a) focuses on an short interval (five main cavity lifetimes) from a sample realization of the cavity photon number $\langle a^{\dagger}a(t) \rangle_{\rm REC}$ obtained via a direct photodetection unraveling of ME~\eqref{eq:MEJC}. The quantum trajectory average transitions from the bright (B) to the dim (D) state in the course of about the average photon lifetime $(2\kappa)^{-1}$. A small fraction of that time is occupied by the coherent-state localization between the B and U states, completed within $\kappa \Delta t_{\rm loc} \approx 0.106$; establishing a lower bound, Eq.~\eqref{eq:Dtloc} (with $\alpha_1$ and $\alpha_2$ taking the values identified by the peak locations of the conditioned $Q$ function) gives $\kappa \Delta t_{\rm loc} \approx 0.097$. 

In parallel, the main and auxiliary scaled cavity amplitudes are shown in the insets. The scaling factor $\kappa^{\prime}/g^{\prime}$ for $\varepsilon^{\prime}=0$ confirms the validity of the adiabatic elimination [see Eq.~\eqref{eq:ava}]. Right underneath, in frame (b), we observe the simultaneous response of the auxiliary cavity which is effectively driven by the stochastic field $\alpha_t$. When the JC system is in the bright or in the dim state, the scaled auxiliary cavity output flux and transmission oscillate about $|\tilde{\alpha}_1|^2/4$, corresponding to the external field drive amplitude. Large deviations from the latter value with distinct frequencies are noted in between these two states, however, starting once we get past the dip due to the coherent cancellation of the effective auxiliary intracavity field. Evidently, we assume that in this instance the experimenter does not take the decision to cut the circuit by moving the control switch from ${\rm P}_A$ to ${\rm P}_B$. Coherent localization, where the bright and unstable (U) states coexist, lasts about one tenth of the main cavity lifetime, as predicted by Eq.~\eqref{eq:Dtloc}. At the center of this continuous evolution lies the time when the transmission dip is attained. At that time, we find an equal-weight superposition of the bright and unstable state amplitudes in the conditioned $Q$ function and cavity density matrix, as displayed in the three insets of Fig.~\ref{fig:FIG2}(b).

In fact, the coherent-state superposition between the bright and unstable states remains in place even past the coherent localization process (at the time $\kappa t_1=6.650$); the conditioned cavity matrix shown in inset (ii) of Fig.~\ref{fig:FIG2}(c) reveals ostensible off-diagonal elements, distinguishing a pure from a mixed state, in spite of the very low occupancy of the bright state. The conditioned photon occupation probabilities $p_n(t_1)=[\rho_{c}(t_{1})]_{nn}$ corresponding to the bright-state peak fall below 0.2\%, while ${\rm Im}\{[\rho_{c}(t_{1})]_{mn}\} \sim 0.01$ at the off-diagonal distributions (for a comparison with analytical results see Ref.~\cite{JumpsBist}). In contrast, the two peaks in the $Q$ function representing the steady-state solution of the ME~\eqref{eq:MEJC} and pictured in the inset (i) of Fig.~\ref{fig:FIG2}(d), correspond to a statistical mixture of the B and D states. 

The persisting coherence in single realizations prompts us to look at the variation of the scaled field $x^{\prime}$ and $y^{\prime}$ quadratures in the auxiliary cavity [last two insets of Fig.~\ref{fig:FIG2}(d)]. Their derivatives attain a clear minimum and a maximum, respectively, midway through the coherent localization. Complementing the criterion based on the dip in transmission, one could employ the extrema of these quantities to ascertain an equal-weight coherent superposition. When doing so, in inset (i) of Fig.~\ref{fig:FIG2}(d) we note that the corresponding times are very close to $t_{\rm dip}$ but do not quite coincide. This instance heralds the presence of an uncertainty associated with the desired conditioned state determination, a theme which will occupy our attention in the remainder of this report.

\section{Complementary methods of reading a macroscopic quantum jump}
\label{sec:compmethods}

We are ready to proceed to \underline{Stage 2} of the proposed experiment. It involves mode-matched heterodyne/homodyne detection on the main---at this stage empty---cavity after having set $g=g^{\prime}=\varepsilon=0$, an event simultaneous to moving the control switch from $\mathrm{P}_A$ to $\mathrm{P}_B$ and resetting the time origin $t_{\rm dip}\to t=0$. The main cavity field at that time is in a pure state, namely the coherent-state superposition of the form
\begin{equation}\label{eq:psi01}
|\psi_{\rm REC}(t_{\rm dip})\rangle \to |\psi(0)\rangle=c_1 |\alpha_1 \rangle + c_2 |\alpha_2 \rangle,
\end{equation}
where $|\alpha_1 \rangle$ and $|\alpha_2 \rangle$ are two macroscopically distinct coherent states, which can be identified from the main cavity field distribution conditioned on a particular downward switch along a trajectory. The state is consistently normalized since it arises from a trajectory unraveling a probability density operator with a given record probability~\cite{CarmichaelBook2}. The two coefficients satisfy $|c_1|^2 + |c_2|^2=1$, in light of the cross terms vanishing as $\exp(-|\alpha_1-\alpha_2|^2) \to 0$. We are thus concerned with the very decay mechanism of macroscopic superposition states~\cite{Walls1985,Phoenix1990,Kim1992,Brune1992,Zurek2003,RomeroIsart2010,Carmichael2013Ch4, Girvin2019,Dakic2017,Qin2019,Qin2021}, those conditionally produced by metastable state changes in quantum amplitude bistability. The number of photons involved in such macroscopic changes scales with $n_{\rm sat}$~\cite{Carmichael2015}.

The QND protocol we are here implementing directly probes the state amplitudes $\alpha_1, \alpha_2$, which are both exponentially decaying functions of time during the interval from $t_{\rm dip}-\Delta t_{\rm loc}/2$ to $t_{\rm dip}$ (during Stage 1), nevertheless taken as constants over the short ($\ll \kappa^{-1}$) timeslace considered ($e^{-\kappa \Delta t_{\rm loc}/2} \approx 1$). Also, somewhat more indirectly, the protocol also measures the complex coefficients $c_1, c_2$  which are rapidly varying functions of time over the abovementioned interval $\Delta t_{\rm loc}/2$, prior to the decision making (see Sec.~\ref{subsec:PSc}). These two coefficients are generally stochastic quantities subject to the uncertainty attributed to the time instance when the decision is made to move the control switch from ${\rm P_A}$ to ${\rm P_B}$ over different realizations. In the ideal case, their magnitudes $|c_1|, |c_2|$ follow a narrow  {\it a posteriori} distribution about $1/\sqrt{2}$. This instance justifies the ``midway through the coherent localization'' qualification, since, to very good accuracy, the conditioned cavity photon number at $t_{\rm dip}$ is $\langle a^{\dagger}a(t_{\rm dip})\rangle=\tfrac{1}{2}(|\alpha_1|^2 + |\alpha_2|^2)$; after the lapse of $\Delta t_{\rm loc}$, the macroscopic switch is considered to have concluded, with the cavity photon number dropping to $|\alpha_2|^2$~\cite{JumpsBist}. In principle, the conditioned amplitudes $\alpha_1$ and $\alpha_2$ are also stochastic quantities, reflecting the relatively weak fluctuations of the bright state and, more importantly, the intense fluctuations of the unstable state. 

Equation~\eqref{eq:psi01} connects the two principal stages by providing the initial state to the empty cavity mode. Its free decay generates a stochastic photocurrent record which we aim to construct, once again on the basis of continuously monitoring the open system output field. 

For that purpose, we need to introduce a local oscillator (LO) as a coherent field with high photon flux added to the field to be measured---the main cavity source field. Continuous photoelectric detection is now performed on their superposition. To be more precise, in the configuration depicted in Fig.~\ref{fig:FIG1}, the charge~\footnote{When using the term {\it charge} we have in mind a measured signal produced by a generalized detector, such as those employed in current circuit QED architectures.} arises from the detection of two superimposed fields with a $\pi$ phase differnece between the superpositions~\cite{CarmichaelBook2}. The scheme uses interference to uncover aspects of scattering associated with a wave amplitude and a spectrum. Homodyne detection~\cite{Carmichael1993QTII, Wiseman1993, Welsch1999, CarmichaelBook2} sets a clear phase reference provided by the LO phase $\theta$; the LO frequency matches $\omega_0$ of the cavity mode. In contrast, in heterodyne detection we detune the LO (frequency $\omega_{\rm LO}$) from the cavity field by an amount $\Delta\omega_{\rm LO}\equiv \omega_{\rm LO}-\omega_0,\quad |\Delta \omega_{\rm LO}| \gg \kappa$, which amounts to the replacement of $\theta$ by the time-varying phase factor $-\Delta \omega_{\rm LO} t$. In connecting the two unravelings, it is useful to note that heterodyne detection is equivalent to running $x \equiv \mathcal{A}_0$ and $y \equiv \mathcal{A}_{\pi/2}$ quadrature homodyne measurements in parallel. To accomplish this unraveling scheme, we split the signal beam at a $50/50$ beam splitter to measure $x$ on one beam splitter output, and $y$ on the other. In the mode-matched variant of both schemes---designed to match the decay of the main cavity mode---the LO amplitude in photon flux units is tailored to the exponentially decaying function of time, $|\varepsilon_{\rm LO}| e^{-\kappa t}$.

\subsection{Heterodyne detection}
\label{subsec:hetdet}

We start with mode-matched heterodyne detection. The stochastic Schr\"{o}dinger equation (SSE) with heterodyne-current records in the interaction picture is 
\begin{equation}
d|\psi_{\rm REC}\rangle=[-\kappa a^{\dagger}a\,dt + (G|\varepsilon_{\rm LO}|e^{-\kappa t})^{-1} \sqrt{2\kappa}a\,dq]|\psi_{\rm REC}\rangle,
\end{equation}
with
\begin{equation}
dq=G|\varepsilon_{\rm LO}|e^{-\kappa t} \left(\sqrt{2\kappa} \langle a^{\dagger}\rangle_{\rm REC}+ dZ \right).
\end{equation}
In the above, $dZ=(dW_x + i dW_y)/\sqrt{2}$ is a complex-valued Wiener increment with covariances $\overline{dZ dZ}=\overline{dZ^{*} dZ^{*}}=0$, $\overline{dZ dZ^{*}}=dt$; $G$ is a generalized detector gain coefficient.

The integrated (or cumulative) charge
\begin{equation}
Q\equiv \sqrt{2\kappa}(G|\varepsilon_{\rm LO}|)^{-1}\int_0^t dq,
\end{equation}
deposited in the heterodyne detector is governed by the stochastic differential equation~\cite{CarmichaelBook2}
\begin{equation}\label{eq:SDE1}
dQ=-\frac{\partial}{\partial Q^{*}}V(Q,Q^{*},t) (2\kappa e^{-2\kappa t}dt) + \sqrt{2\kappa}\,e^{-\kappa t} dZ,
\end{equation}
where we have introduced a time-varying potential which explicitly depends on the initial state, {\it i.e.} the state of the cavity field when the control switch is moved to the position $\mathrm{P}_B$. The potential is defined as
\begin{equation}
V(Q,Q^{*},t)=-\ln \left[\langle \psi(0)|e^{Q^{*}a^{\dagger}} e^{-2\kappa t} e^{Qa}|\psi(0)\rangle  \right],
\end{equation} 
which, for $|\alpha_1-\alpha_2|^2 \gg 1$, satisfies
\begin{equation}\label{eq:VHet}
\begin{aligned}
e^{-V(Q,Q^{*},t)}\approx \sum_{i=1}^2&|c_i|^2 \exp[-|\alpha_i|^2 (1-e^{-2\kappa t})]\\
&\times\exp(\alpha_i Q + \alpha_i^{*}Q^{*}),
\end{aligned}
\end{equation}
where cross terms are neglected, scaling as  $|\langle \alpha_1|\alpha_2 \rangle|=e^{-\tfrac{1}{2}|\alpha_1-\alpha_2|^2} \ll 1$. 

In the long-time limit $t\to \infty$, the above expression is recast into the form
\begin{equation}\label{eq:Vhet2}
e^{-V(Q,Q^{*},t\to \infty)}\approx \sum_{i=1}^2|c_i|^2 \exp(-|\alpha_i-Q^{*}|^2) \exp(|Q|^2),
\end{equation}
which shows that there are two potential wells centred at the positions $Q=\alpha_1^*$ and $Q=\alpha_2^{*}$ in the complex-$Q$ plane; individual realizations of $Q$ solving Eq.~\eqref{eq:SDE1} will be trapped in either of the two when all light will have left the main cavity. We note that due to the large separation between the two amplitudes, any interference associated with the macroscopic-state superposition is effectively lost in the potential~\eqref{eq:VHet}; so is the relative phase between the coefficients $c_1$ and $c_2$. 

\begin{figure*}
\includegraphics[width=\textwidth]{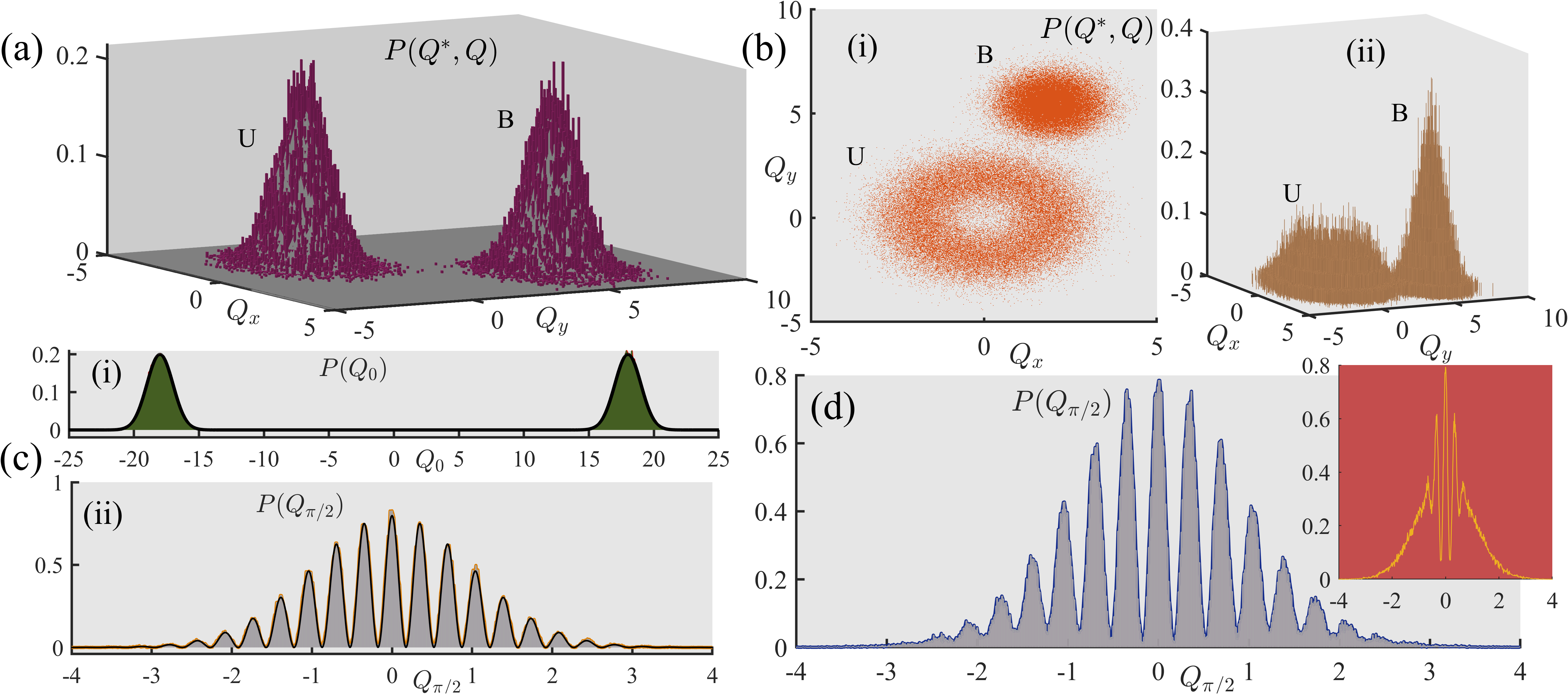}
\caption{{\it Statistical properties of the integrated charge.} {\bf (a)} Two-dimensional histogram of the complex integrated charge $Q_x+iQ_y$ normalized to a probability density function $P(Q^{*},Q)$, collected from mode-matched heterodyne detection on the main cavity output at ten cavity lifetimes after the control switch has been moved to position $P_B$. For the $10^5$ realizations recorded, the initial cavity state is taken as~\eqref{eq:psi01} with $c_1=c_2=1/\sqrt{2}$, while the conditioned amplitudes are assumed identical for every jump across different realizations, equal to $\alpha_1=1.95-5.45i$ (bright state) and $\alpha_2=-1.40+0.85i$ (unstable state), corresponding to the peak positions of the $Q$ function displayed in inset (i) of Fig.~\ref{fig:FIG2}(b). The bars are superimposed on the surface plot of the $Q$ function $\mathcal{Q}(Q_x + iQ_y)=[1/(2\pi)]\{\exp[-(Q_x-1.95)^2-(Q_y-5.45)^2] + \exp[-(Q_x+1.40)^2-(Q_y+0.85)^2]\}$. {\bf (b)} Same principle as in (a) but instead with the unstable state coefficient $|c_2|^2$ assuming random values uniformly distributed in $(0,1)$, while $|c_1|^2=1-|c_2|^2$, and with the unstable state amplitude taking a fixed modulus $|\alpha_2|$ but a uniformly distributed phase in $(0, 2\pi)$. A surface density projection of the histogram is given in (i) and the bar graph is depicted in (ii). {\bf (c)} Histograms normalized to a probability density function of $10^5$ homodyne current records $P(Q_{\theta})$ at $t\to \infty$ ($\eta \to 1$) for the initial state~\eqref{eq:cat} with $\phi_0=0$ and $A=9$. The LO phase is set to $\theta=0$ (i) and $\theta=\pi/2$ (ii). Solid black lines plot the analytical expression for $P(Q_{\theta})$ in Eq.~\eqref{eq:PQan}. {\bf (d)} Same as in (c) for $\theta=\pi/2$ but with the amplitude $A$ exhibiting fluctuations across different trajectories, which are Gaussian-distributed with mean $\bar{A}=9$ and standard deviation $\sigma=0.1\sqrt{\bar{A}}$ (main frame) and  $\sigma=0.5\sqrt{\bar{A}}$ (inset). B and U stand for the bright and unstable states, respectively.}
\label{fig:FIG3}
\end{figure*}

Let us deal with the idealized scenario where the conditioned amplitudes $\alpha_1$ and $\alpha_2$, as well as the weights $c_1$ and $c_2$, are deterministic and invariant across quantum jumps in different realizations. As we may have anticipated from the term $\exp(-|\alpha_i-Q^{*}|^2)$ in Eq.~\eqref{eq:Vhet2}, the long-time distribution $P(Q, Q^{*}, t\to \infty)$ of the integrated charge yields the $Q$ function of the {\it initial} cavity field prior to the free decay, $\mathcal{Q}_{t=0}$, namely
\begin{equation}
\begin{aligned}
P(Q, Q^{*}, t\to \infty)&=\pi^{-1} \langle \psi(0)|Q^{*}\rangle \langle Q^{*}|\psi(0)\rangle\\
&=\mathcal{Q}_{t=0}(Q^{*},Q),
\end{aligned}
\end{equation}
where $|Q^{*}\rangle=e^{-\frac{1}{2}|Q|^2} e^{Q^{*}a^{\dagger}}|0\rangle$ stands for the coherent state of complex amplitude $Q^{*}$. Such is the case for the complex charge distribution depicted in Fig.~\ref{fig:FIG3}(a); the 2D histogram of the integrated charge $Q$ deposited in the heterodyne detector after all photons have left the main cavity is well approximated by the $Q$ function $\mathcal{Q}(Q_x + iQ_y)=[1/(2\pi)]\{\exp[-(Q_x-1.95)^2-(Q_y-5.45)^2] + \exp[-(Q_x+1.40)^2-(Q_y+0.85)^2]\}$---a sum of two equally weighted Gaussian distributions. The two Gaussian functions are peaked at the location of the (complex conjugate of the) conditioned amplitudes $\alpha_1^{*}=1.95+5.45i$ and $\alpha_2^{*}=-1.40-0.85i$, as read from the peak positions of the conditional {\it quasi}probability distribution of the main cavity field $\mathcal{Q}(x+iy;t_{\rm dip})$, recorded at the auxiliary cavity transmission dip. Here we neglect squeezing, a direct consequence of the JC nonlinearity~\cite{CarmichaelBook2}, and attribute the peak positions to coherent states with the corresponding complex amplitudes. The identification is justified by the localization process between coherent states, which models the JC downward jump times and cavity distributions to a very good accuracy~\cite{JumpsBist}.  

The stochastic nature of a jump detection and its correlated quantum state~\eqref{eq:psi01} enters in Fig.~\ref{fig:FIG3}(b). It is inherently linked to the decision on whether and when exactly to move the control switch and initiate \underline{Stage 2}. We thus consider two extremes: that there is a time uncertainty of order $\sim \Delta t_{\rm loc}$ in moving the control switch and that, simultaneously, the unstable state location fluctuates over a ring in phase space, as suggested by the $Q$ functions of Fig.~\ref{fig:FIG2}(c). The uncertainty in time is accounted for by uniformly distributed weights $|c_1|^2$, $|c_2|^2$ in Eq.~\eqref{eq:Vhet2}, while the uncertainty in phase space is accounted for by a uniformly distributed phase of the conditioned unstable-state amplitude $\alpha_2$~\footnote{Strictly speaking, the time uncertainty in moving the control switch during the localization is translated to a variation in the exponentially decaying amplitudes $\alpha_1, \alpha_2$ over $\kappa \Delta t_{\rm loc}$  (see Sec.~\ref{subsec:PSc}). This variation, however, which is here altogether neglected, may well prove indistinguishable from the inherent quantum fluctuations in the bright and unstable state amplitudes.}. The result is a ``continuously rotated'' distribution about the origin of the complex $Q$ plane, corresponding to the unstable state, while the peak at $\alpha_1^{*}$ (conjugate of the bright-state amplitude) remains in place. We infer that the heterodyne current record is able to capture every phase-space ramification of the underlying state evolution as the downward switch unfolds. Even rarely visited regions in phase space, correlated with the control switch movement, will be eventually represented in the long-time histogram of the integrated charge. There is still a crucial piece of the narrative missing though.

\subsection{Homodyne detection}
\label{subsec:homdet}

Unlike its heterodyne counterpart, the integrated charge produced in mode-matched homodyne detection {\it does} reflect the quantum interference between the two macroscopic states in superposition~\cite{YurkeStoler1986, CarmichaelSG1994}. Indeed, we expect that the phase reference originating from the strong LO field, interfering with the cavity field in a coherent-state superposition, will markedly influence the formation of a measured interference pattern. To prove this assertion, we tune the LO to the cavity-mode frequency $\omega_0$. The LO complex amplitude has the same exponentially decaying profile as before, but now a well defined phase $\theta$; we write $\varepsilon_{\rm LO}=|\varepsilon_{\rm LO}| e^{i\theta} e^{-\kappa t}$. 

Consider here a Schr\"{o}dinger cat state~\cite{Dodonov1974, Agarwal1997, HarocheBook, Hacker2019} written in the form
\begin{equation}\label{eq:cat}
|\psi(0) \rangle = \frac{1}{\sqrt{2}} (|A \rangle + e^{i\phi_0}|-A\rangle), 
\end{equation}
with $A=A(\alpha_1, \alpha_2)$ a real number related to the separation of the bright and unstable state amplitudes. The above expression results from the coherent state superposition of Eq.~\eqref{eq:psi01}, taken exactly midway through the localization (with $|c_1|^2=|c_2|^2=\tfrac{1}{2}$), by applying a suitable coherent offset, followed by a rotation of the co-ordinate axes in phase space to match the argument of $\alpha_1-\alpha_2$; in the ideal case, the latter operations result in a deterministic value of $A$. In practice, however, when assessing an ensemble the ensemble of several realizations, the fluctuating amplitudes $\alpha_1, \alpha_2$ do not remain collinear across the different coherent localization events occurring at random times. It proves then convenient to select the phase of the drive field $\varepsilon$ such that $\tilde{\alpha}_1$ lies on the real axis; we then project the conditioned amplitudes onto the real axis, reducing the fluctuations of $\alpha_2$ to the semiclassical estimate $|\tilde{\alpha}_2|$. For a very large separation between the two conditioned states in phase space, $|\alpha_1-\alpha_2|^2 \gg 1$ (such as the one considered in Sec.~\ref{subsec:hetdet}), a coherent offset $-\alpha_1/2$ leaves us to a good approximation with $A=|\alpha_1-\alpha_2|/2 \approx \alpha_1/2 \gg 1$ at zero order, translating the unstable-state fluctuations to small-amplitude (first-order) fluctuations in the random variable $A$. We can restrict our attention to very large amplitudes $A$ generated in the bistable trajectories of the strong-coupling ``thermodynamic limit'' $n_{\rm sat} \to \infty$~\cite{Carmichael2015}, and entailing equally sharp dips in the auxiliary cavity transmission and output flux. In principle, an unequivocal decision on the coherent cancellation produces in turn an equal-weight superposition between the two coherent states in Eq.~\eqref{eq:cat}; fluctuations in the conditioned value of $A$ between different jumps are nevertheless to be retained, and will be discussed further in Sec.~\ref{sec:statdev}.  

The SSE dictating the production of homodyne-current records in the interaction picture explicitly includes the phase reference $\theta$ as~\cite{CarmichaelBook2}
\begin{equation}\label{eq:SSE}
\begin{aligned}
d|\psi_{\rm REC}\rangle&=[-\kappa a^{\dagger}a\,dt \\
&+ (G|\varepsilon_{\rm LO}|e^{-\kappa t})^{-1} e^{-i\theta} \sqrt{2\kappa}a\,dq_{\theta}]|\psi_{\rm REC}\rangle,
\end{aligned}
\end{equation}
where 
\begin{equation}
dq_{\theta}=(G|\varepsilon_{\rm LO}|e^{-\kappa t})\left(2\sqrt{2\kappa}\langle \mathcal{A}_{\theta} \rangle_{\rm REC} + dW \right)
\end{equation}
is the real charge deposited in the detector circuit; $dW$ is a real Wiener increment, Gaussian-distributed with zero mean and variance $dt$. The relative amplitude ($\theta=0$) or phase ($\theta=\pi/2$) in the two components of the conditioned state $|\psi_{\rm REC}\rangle$ diffuses in parallel with the photocurrent. 

Similar to the stochastic differential equation~\eqref{eq:SDE1} of Sec.~\ref{subsec:hetdet}, the integrated charge
\begin{equation}
Q_{\theta} \equiv \sqrt{2\kappa}(G|\varepsilon_{\rm LO}|)^{-1}\int_0^t dq_{\theta}
\end{equation}
obeys~\cite{CarmichaelSG1994}
\begin{equation}\label{eq:SDE2}
dQ_{\theta}=-\frac{\partial}{\partial Q_{\theta}} V(Q_{\theta},\eta) d\eta + d\zeta,
\end{equation}
with $\eta \equiv 1-e^{-2\kappa t}$~\cite{CarmichaelBook2} taking values in the interval $[0,1)$, and $d\zeta$ a Wiener increment with $\overline{d\zeta^2}=d\eta$. The potential, however, has a crucial difference, brought about by the constant phase reference of homodyne detection:
\begin{equation}\label{eq:Vhom}
\begin{aligned}
e^{-V(Q_{\theta},\eta)}&=\cosh(2Q_{\theta}A\cos\theta)\exp[A^2(1-2\eta\cos^2\theta)]\\
& + \cos(\phi_0+2 Q_{\theta}A \sin\theta)\exp[-A^2(1-2\eta\sin^2\theta)].
\end{aligned}
\end{equation}
The second term in Eq.~\eqref{eq:Vhom} explicitly uncovers quantum interference, while the form of $V(Q_{\theta},\eta)$ qualitatively changes with the LO phase $\theta$~\cite{Carmichael1999}. If we neglect fluctuations in the real amplitude $A$ across quantum jumps occurring along different trajectories, the steady-state distribution of $Q_{\theta}$, denoted by $P(Q_{\theta},\eta\to 1)$, measures a marginal of the Wigner distribution representing the state of the main cavity field immediately prior to the period of free decay~\cite{CarmichaelSG1994}. For the cat state of Eq.~\eqref{eq:cat}, one obtains the long-time charge distribution~\cite{CatWPD}
\begin{equation}\label{eq:PQan}
\begin{aligned}
P(Q_{\theta},&\eta\to 1)=[2\sqrt{2\pi}\,\cosh(A^2)]^{-1}e^{-Q_{\theta}^2/2}\\
&\times\{\cosh(2Q_{\theta}A\cos\theta)\exp[A^2(1-2\cos^2\theta)]\\
& + \cos(\phi_0+2 Q_{\theta}A \sin\theta)\exp[-A^2(1-2\sin^2\theta)]\}.
\end{aligned}
\end{equation}
The last term evinces the familiar fringes due to quantum interference, manifested most strongly for $\theta=\pi/2$. At the latter setting, the fringe spacing in the cumulative charge distribution $\propto e^{-Q^2/2}\cos^2(QA)$ is set by the conditioned bright and unstable state separation $A=|\alpha_1-\alpha_2|/2$, as $\Delta Q_{\pi/2}=\pi/A$. 

We appeal anew to the idealized scenario seeing a deterministic amplitude $A$, the same across the ensemble of realizations in hand. Figure~\ref{fig:FIG3}(c) shows that the statistical behaviour of the homodyne-current record is an explicit function of the LO phase. Measuring ``position'' with $\theta=0$ [inset (i)], we obtain two Gaussian distributions centered at $Q_0=\pm 2A$, a fact which may be used to {\it a posteriori} verify the value of $A$ previously identified from the conditioned $Q$ function of the cavity state. Once the value of $A$ has been cross-checked from an ensemble of ``position'' records, we can proceed to set $\theta=\pi/2$ measuring ``momentum'' [inset (ii)], in order to generate a new ensemble and assess its statistics. The obtained histogram demonstrates that a $\cos^2(Q_{\pi/2} A)$ oscillating profile is screened by the Gaussian envelope $e^{-(1/2)\,Q_{\pi/2}^2}$.

Mode-matched homodyne detection also proves able to resolve the phase difference between the two components in the superposition~\eqref{eq:cat}: if $\phi_0$ consistently assumed the value of $\pi$ instead of $0$ across the ensemble, the charge distribution would still be symmetric with respect to the $Q_{\rm \pi/2}=0$ axis, but with a dip at the origin in place of a peak~\cite{YurkeStoler1986, CatWPD}. For other values of $\phi_0$, this mirror symmetry is lost.

\section{Jumps-ensemble statistics and fringe visibility in the charge distribution} 
\label{sec:statdev} 

Armed with these observations, we remain with the homodyne scheme detailed in Sec.~\ref{subsec:homdet} to take into consideration the fluctuations of $A$ across different jumps, in what leads to the collected ensemble of initial states, which are read midway through the coherent localization. These are due to two main sources: {\bf (a)} the temporal fluctuations of the conditioned amplitude $\alpha_1$ when the JC system is in the bright metastable state preceding the coherent-state localization. Quantum trajectories, such as the one depicted in Fig.~\ref{fig:FIG2}(a), reveal that the extent of these fluctuations is significantly lower than the standard deviation $|\tilde{\alpha}_1|$ of a Poisson distribution with mean $|\tilde{\alpha}_1|^2$, taken as reasonable candidate; {\bf (b)} the aforementioned unstable-state fluctuations, like those depicted in Fig.~\ref{fig:FIG2}(c). The latter are relatively intense with respect to $|\tilde{\alpha}_2|$ but rather small in size when their square is compared to the amplitude separation $|\tilde{\alpha}_1-\tilde{\alpha}_2|^2$. Consequently, as we discussed above, they can be projected in phase space to the line joining the origin with $\tilde{\alpha}_1$, while the latter can be taken real after a suitable selection of the drive amplitude $\varepsilon$ in Eq.~\eqref{eq:MEfull}. For the two conditioned state amplitudes, identified by the $Q$ function of Fig.~\ref{fig:FIG2}(b) [inset (i)] and displayed as charge distribution peaks in Fig.~\ref{fig:FIG3}(a), we obtain $|\alpha_2/(\alpha_1-\alpha_2)|^2 \approx 0.053$, justifying our approximation at the level of individual realizations.

Both sources of fluctuations may be combined and phenomenologically modeled by a Gaussian function of mean $\bar{A} \approx |\tilde{\alpha}_1-\tilde{\alpha}_2|/2$ and a standard deviation $\sigma/\sqrt{\bar{A}} \lesssim 0.1$ consistent with the first-order corrections to the mean. Hence, the amplitude $A$ in Eq.~\eqref{eq:cat} entering the potential~\eqref{eq:Vhom} through the initial state, is a random variable with the aforementioned Gaussian distribution. For $\sigma/\sqrt{\bar{A}}=0.1$, the distribution $P(Q_{\pi/2})$ remains virtually intact apart from the partial erasure of some ``further out'' fringes screened by the Gaussian envelope, as we can see in the main frame of Fig.~\ref{fig:FIG3}(d). Even in the extreme case of $\sigma/\sqrt{\bar{A}}=0.5$, one can still read the interference pattern out of the ensemble fluctuations; the ``closest-to-center'' fringes are in place, separated by $\pi/\bar{A}$ from the central peak.

A qualitatively similar degradation of the fringe visibility is noted in Fig. 2 of~\cite{YurkeStoler1986} when the detector efficiency in measuring a coherent-state superposition via homodyne detection decreases. In our case, however, the dephasing over an {\it ensemble of quantum trajectories} is not caused by any device inefficiency or loss, but rather by inherent quantum fluctuations in the main cavity field---the source field---reflected in the distribution of the random variable $A(\alpha_1, \alpha_2)$. We remark that a further visibility degradation to the one discussed above and observed in Fig.~\ref{fig:FIG3}(d) would reflect a varying phase $\phi_0$ between the two components in Eq.~\eqref{eq:psi01} across different jumps in the obtained ensemble. As we saw in Sec.~\ref{subsec:hetdet}, this relative phase does not have operational consequences in heterodyne detection. 

\subsection{Photon scattering from a decaying coherent-state superposition}
\label{subsec:PSc}

To understand a primary source of the above deviations, as seen from the quantum-trajectory perspective, we can produce an estimate of the probability that a coherent localization process concludes to the unstable state amplitude $\alpha_2$. While the system is still in the bright metastable state, our starting point is determined by the emergence of the unstable state during continuous monitoring, leading to a conditioned state superposition of the form $ \mathrm{c}_1 |\alpha_1 \rangle + \mathrm{c}_2 |\alpha_2\rangle$, with $|\mathrm{c}_1|^2 > |\mathrm{c}_2|^2$ and $|\mathrm{c}_1|^2 + |\mathrm{c}_2|^2=1$ (we distinguish the coefficients of this state from those of Eq.~\eqref{eq:psi01} with roman font). The said formation heralds a possible downward switch with a set probability. For convenience, we take here this particular time instant as the origin $t=0$, instead of the time origin set by the actuation of Stage 2 in Sec~\ref{sec:compmethods} (the amplitude ratio of $\alpha_{1,2}$, identified via conditioned cavity distributions, in the two cases is $e^{-\kappa \Delta t_{\rm loc}/2} \approx 1$). 

For the purpose of this argument, at $t \geq 0$ we remain in Stage 1 of our protocol throughout the evolution, where the photoelectron ``clicks'' are registered by the APD, but instead consider the free decay of the cavity mode setting $\varepsilon=g=0$~\cite{JumpsBist}. For a negligible overlap $\exp(-|\alpha_1-\alpha_2|^2)$, which suppresses cross-terms alongside the relative-phase effect between the two components, the un-normalized conditioned state at time $t$ is~\cite{Carmichael2013Ch4}
\begin{equation}\label{eq:psirec}
\begin{aligned}
|\psi_{\rm REC}(t)\rangle=&\sum_{i=1,2} \mathrm{c}_i (\sqrt{2\kappa} \alpha_i e^{-\kappa t_n})\ldots (\sqrt{2\kappa} \alpha_i e^{-\kappa t_1})\\
&\times\exp \left[-\tfrac{1}{2}|\alpha_i|^2 (1-e^{-2\kappa t})\right]\,|\alpha_i e^{-\kappa t}\rangle,
\end{aligned}
\end{equation}
where photoelectric emission ``clicks'' are registered at times $t_1, t_2,\ldots t_n$. In what follows we take $t\sim \Delta t_{\rm loc}$. During an individual localization instance, modeled by Eq.~\eqref{eq:psirec} with $n=0$ (a so-called {\it null-measurement record}), the relative weight $r(t)$ of the two components in the probability density corresponding to the superposition (with $r(0)=|\mathrm{c}_1|^2/|\mathrm{c}_2|^2$) reads 
\begin{equation}
\frac{r(t)}{r(0)}=\frac{\exp[-|\alpha_1|^2(1-e^{-2\kappa t})]}{\exp[-|\alpha_2|^2(1-e^{-2\kappa t})]} \approx e^{-2\kappa (|\alpha_1|^2-|\alpha_2|^2) t}.
\end{equation}
Evaluated at $t=\Delta t_{\rm loc}/2$, the weight $r$ corresponds to the ratio $|c_1/c_2|^2$ in Eq.~\eqref{eq:psi01}, emerging as a result of the decision to actuate Stage 2. Specifically, setting $r(\Delta t_{\rm loc}/2)=1$ to realize an equal-weight superposition --- akin to the one ``midway through the continuous macroscopic switch'' in Sec~\ref{sec:compmethods} --- connects the two wavefunctions by providing an estimate on the initial relative weight $r(0)$, particular to the null-measurement record. 

By integrating the record probability density $\langle \psi_{\rm REC}(t)|\psi_{\rm REC}(t)\rangle$ over all possible count times, under the idealized assumption that $\alpha_1$ and $\alpha_2$ remain the same across all jumps, we obtain the count distribution as a weighted sum of two Poisson distributions
\begin{equation}
\begin{aligned}
P(n,t)= \sum_{i=1,2}& |\mathrm{c}_i|^2\frac{\left[|\alpha_i|^2(1-e^{-2\kappa t})\right]^n}{n!}\\
&\times\exp\left[-|\alpha_i|^2(1-e^{-2\kappa t})\right].
\end{aligned}
\end{equation}
In view of the systematic uncertainty in the time of transition between the two principal stages, we take $t \sim \Delta t_{\rm loc}$. Setting then $t=\Delta t_{\rm loc}$ [taken from Eq.~\eqref{eq:Dtloc} with $\tilde{\alpha}_i \to \alpha_i$] for a given $n$, yields photon count probabilities as functions of $\alpha_1, \alpha_2$, and $|\mathrm{c}_2|^2$. For $n=0$ (no counts) we therefore obtain the probability $p_0\equiv P(0, \Delta t_{\rm loc})$ that the coherent localization reaches {\it uninterrupted} the unstable state, retaining its inversion symmetry about mid-point~\cite{JumpsBist}. The continuous-evolution (from bright to unstable state) probability --- that of a macroscopic downward switch with mininal duration --- is approximated as
\begin{equation}\label{eq:p0}
\begin{aligned}
p_0\approx & (1-|\mathrm{c}_2|^2) \exp(-2\kappa |\alpha_1|^2 \Delta t_{\rm loc})\\
&+ |\mathrm{c}_2|^2 \exp(-2\kappa |\alpha_2|^2 \Delta t_{\rm loc}),
\end{aligned}
\end{equation}
which shows that the unstable-state weight in the initial ($t=0$) superposition is translated into the likelihood of a macroscopic switch. We note that even though typically $|\alpha_1|^2 \gg |\alpha_2|^2$ at $t=0$, we cannot approximate $p_0$ by the second term  in Eq.~\eqref{eq:p0}, since the two terms are of the same order of magnitude when $|\mathrm{c}_2|^2 \lesssim 0.1$. 

Setting now $n=1$ gives the probability $p_1\equiv P(1, \Delta t_{\rm loc})$ of one photoelectric emission working against coherent localization to the unstable state (with amplitude $\alpha_2 e^{-\kappa \Delta t_{\rm loc}}$). This quantity is approximated as
\begin{equation}\label{eq:p1}
\begin{aligned}
p_1\approx & (1-|\mathrm{c}_2|^2) (2\kappa |\alpha_1|^2 \Delta t_{\rm loc}) \exp(-2\kappa |\alpha_1|^2 \Delta t_{\rm loc})\\
&+ |\mathrm{c}_2|^2 (2\kappa |\alpha_2|^2 \Delta t_{\rm loc}) \exp(-2\kappa |\alpha_2|^2 \Delta t_{\rm loc}),
\end{aligned}
\end{equation}
where the relative weight between the two terms is enhanced by $|\alpha_1/\alpha_2|^2 \gg 1$ in comparison to $p_0$. Each of these first-photon emission events increases the conditioned cavity photon number average in {\it individual} realizations, since the relative weight between the two components of the superposition is updated by a factor $|\alpha_1/\alpha_2|>1$ --- reflected in Eq.~\eqref{eq:p1} as an ensemble average. More importantly, it conditions a new starting point for a coherent-state superposition with different initial amplitudes. The conditioned cavity state thus contributes to the deterioration of the eventual charge fringe visibility in case it provides the initial state to the free decay in Stage 2.  

For a fixed bright-state amplitude $|\alpha_1|$ in the regime of bistability considered in the figures of this report, $p_0 \lesssim 0.1$ can be considered as a function of $|\alpha_2| \gtrsim 1$, whose maximum and monotonicity depend on the weight $|\mathrm{c}_2|^2$ as a parameter.  At the same time, we find that the probability of a single count $p_1$ exceeds $p_0$ when $|\mathrm{c}_2|^2 \ll 1$, as does the two-counts probability $p_2>p_1$.  With a few further ``clicks''  registered along the way ($n \gtrsim 2$), coherent-state localization in the course of an individual realization is typically concluded~\footnote{The photoelectron count ``clicks'' registered past the localization time $\Delta t_{\rm loc}$ typically correspond to the free decay of the eventually established coherent state, with amplitude either $\alpha_1$ or $\alpha_2$ (without taking into account their small damping), and are to be excluded from consideration. We emphasize that the mapping of the JC switch to the empty-cavity damping is valid only during intervals $\sim \Delta t_{\rm loc}$.} with the cavity state most often reverting to the bright, {\it i.e.}, the macroscopic downward switch is aborted. 

\begin{figure*}
\includegraphics[width=\textwidth]{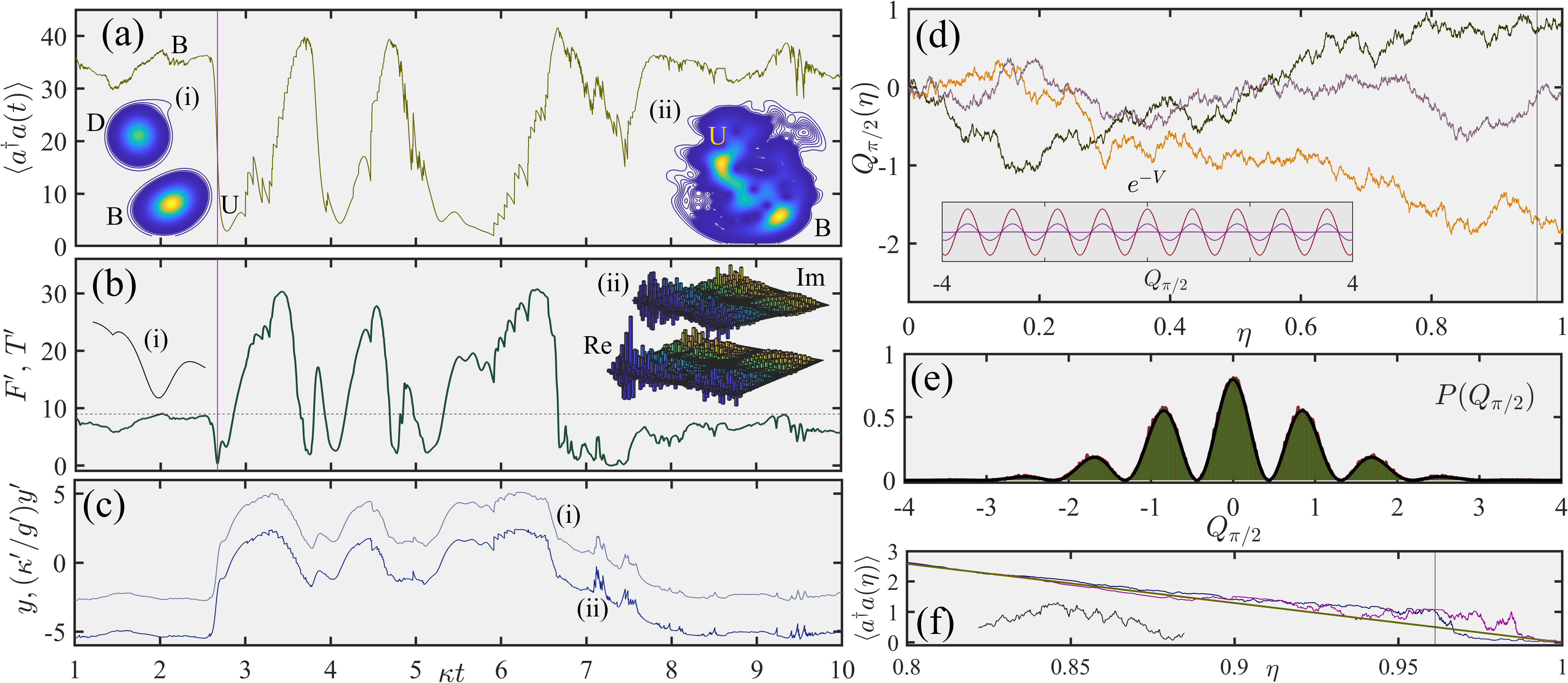}
\caption{{\it Fluctuating vs. metastable states}. {\bf (a)} Sample trajectory interval of the main cavity photon number $\langle a^{\dagger}a(t)\rangle=\langle a^{\dagger}a(t)\rangle_{\rm REC}$ against nine cavity lifetimes, including a downward quantum jump from the bright state which is then followed by large-photon fluctuations. Inset (i) depicts a schematic contour plot of the steady-state main cavity $Q$ function $\mathcal{Q}_{\rm ss}(x+iy)$, while inset (ii) shows the conditioned $Q$ function of the main cavity field $\mathcal{Q}(x+iy; t_{\rm dip})$ at the time $t_{\rm dip}$ indicated by the vertical purple line. {\bf (b)} Scaled outgoing photon flux and transmission from the auxiliary cavity, as defined in Eq.~\eqref{eq:TF}, against the same dimensionless time interval focused upon in (a). The external field amplitude is set to $-\tilde{\alpha}_1/2$, where $\tilde{\alpha}_1\approx 2.448 -5.445i$ solves Eq.~\eqref{eq:neocl}. The purple line in frames (a, b) indicates the position of the dip in flux and transmission, occurring at $\kappa t_{\rm dip}=2.668$. The horizontal line marks the position of $|\tilde{\alpha}_1|^2/4 \approx 8.957$. Inset (i) focuses on the asymmetry of the dip over $0.2$ cavity lifetimes, and inset (ii) depicts barplots of the conditioned main cavity density matrix $[\rho_{c}(t_{\rm dip})]_{mn}$ (real and imaginary parts indicated accordingly). {\bf (c)} $y$-quadrature of the main cavity field against the same time interval as in (a,b) and scaled quadrature $(\kappa^{\prime}/g^{\prime}) y^{\prime}$ of the auxiliary cavity field, plotted by curves (i) and (ii), respectively. {\bf (d)} Three sample realizations against the scaled time $\eta \equiv 1-e^{-2\kappa t}$ of the integrated charge $Q_{\pi/2}$ produced as homodyne current records solving Eq.~\eqref{eq:SDE2}. The coherent offset is set to the value $(\alpha_1 + \alpha_2)/2$, with $\alpha_1 \approx 2.95-5.35i$ (bright state) and $\alpha_2 \approx -2.05-0.20i$ corresponding to the peak locations of the conditioned $Q$ function in [(a), inset (ii)]. The LO phase is set to $\theta=\pi/2$ relative to the axis joining $\alpha_1$ and $\alpha_2$. The initial state is set to that of Eq.~\eqref{eq:cat} with $A(\alpha_1,\alpha_2)=(1/2)|\alpha_1-\alpha_2|\approx 3.59$. The inset depicts $e^{-V}$ vs. $Q_{\pi/2}$ [see Eq.~\eqref{eq:Vhom}], where $V\equiv V(Q_{\pi/2}; \eta)$, at the progressing times $\eta=0.8, 0.95$ and $0.99$ corresponding to the functions oscillating with increasing amplitude. {\bf (e)} Histogram of $Q_{\pi/2}$ from an ensemble of $10^5$ realizations solving Eq.~\eqref{eq:SDE2} in the limit $\eta \to 1$ under the same measurement procedure followed in (d). The analytical solution of a Fokker--Planck equation [Eq.~\eqref{eq:PQan}] for $\theta=\pi/2$ and $A=3.59$ is plotted in thick black line. {\bf (f)} Three sample trajectories of the empty main cavity photon number $\langle a^{\dagger}a(\eta)\rangle$ when its output field is measured with homodyne detection. The LO phase is set to $0$ (single straight decaying line) and $\pi/2$ (pair of non-monotonic decay profiles). Vertical lines in frames (d) and (f) indicate the time $\eta_m=1-1/(2A^2) \approx 0.961$. The inset in (f) focuses on a short interval ($\Delta \eta \approx 0.01$) of $\langle a^{\dagger}a(\eta)\rangle$ along a fourth trajectory, past $\eta_m$. B, D, U denote the bright, dim and unstable states, respectively. The steady-state photon number is $\langle a^{\dagger}a \rangle_{\rm ss}\approx 22.83$ [ensemble average after the solution of ME~\eqref{eq:MEJC}]. In the $\mathcal{Q}_{\rm ss}(x+iy)$ of (a), the D state is practically the vacuum while the B state peaks at $\alpha_{1, {\rm ss}}\approx 2.47-5.45i$. The operating parameters read: $g/\kappa=60$, $\varepsilon/\kappa=14.4i$, $\Delta\omega/\kappa=-7.2$, and $\kappa^{\prime}/\kappa=\kappa^{\prime}/g^{\prime}=100$.}
\label{fig:FIG4}
\end{figure*}

\begin{figure*}
\includegraphics[width=0.36\textwidth]{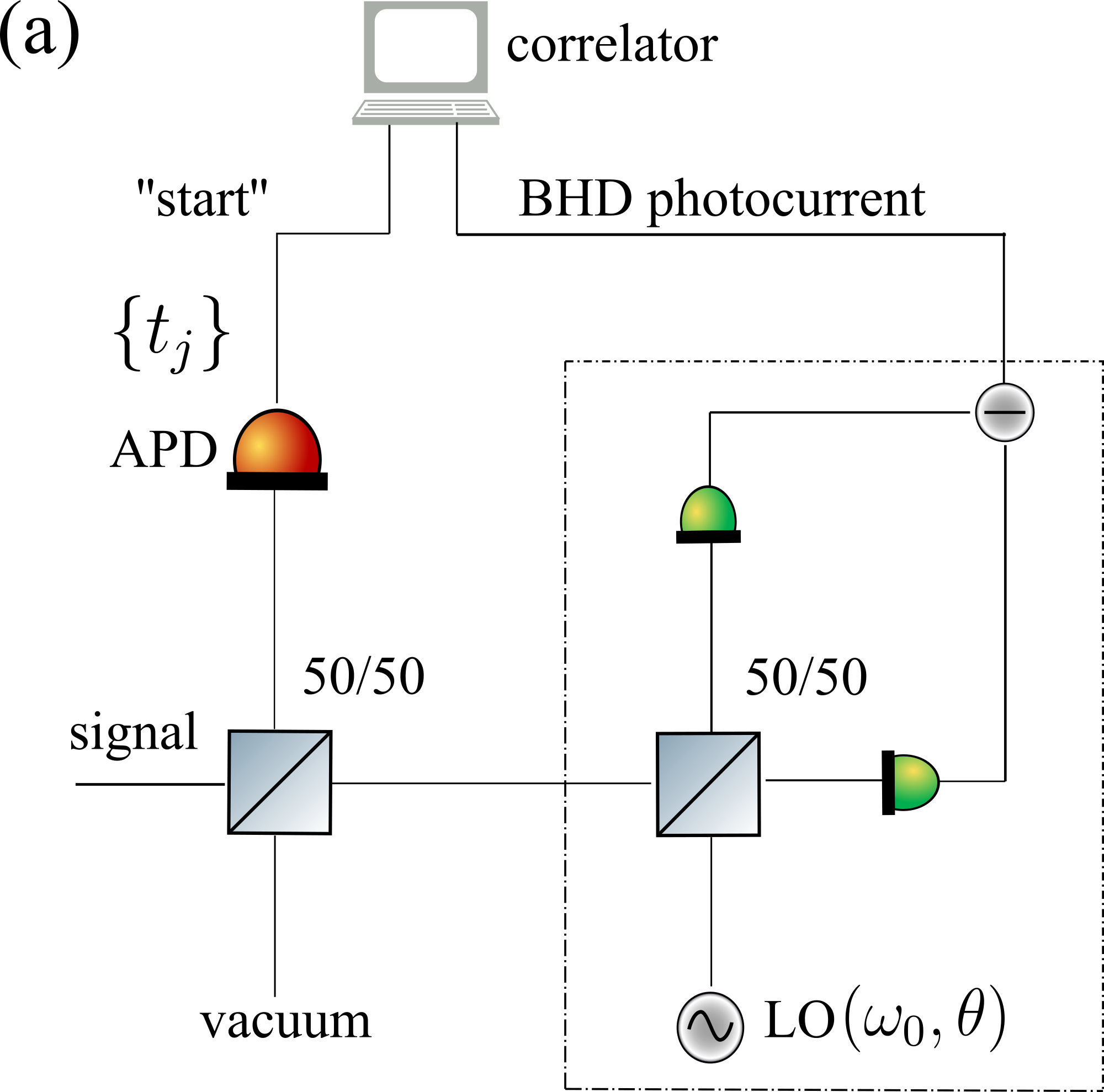}
\includegraphics[width=0.5\textwidth]{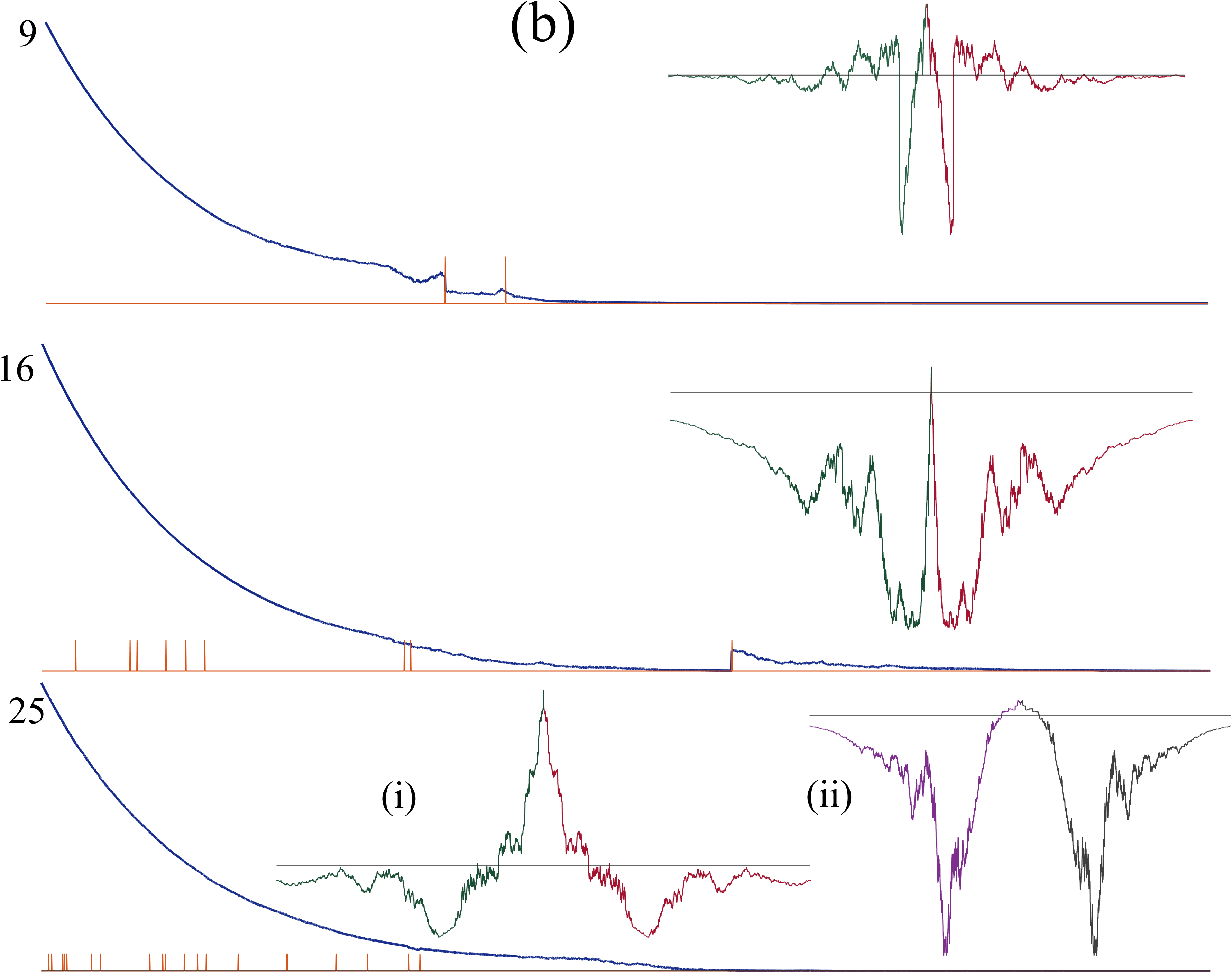}
\caption{{\it Wave/particle correlations read from a macroscopic switch}. {\bf (a)} Schematic illustration of the wave/particle correlator that is to replace the BHD detector of Fig.~\ref{fig:FIG1} when the control switch moves to position ${\rm P_B}$ and the empty main cavity mode decays to the vacuum. Half of the input light is now sent to the BHD detecting the fluctuations of the signal field, while the other half is directed to an APD in the ``start'' channel. The homodyne signal is sampled over a series of $N_s$ time windows, $t_j-\tau_{\rm max} \leq t \leq t_j + \tau_{\rm max}$, each centred on a ``start'' time $t_j$. The LO phase is set to $\theta=\pi/2$ with reference to the position of the conditioned bright and unstable-state amplitudes in phase space. {\bf (b)} Three sample realizations of $\langle a^{\dagger}a (t)\rangle$ against four average cavity lifetimes ($\kappa t_{\rm max}=4$) under the action of the wave/particle correlator of (a). The initial empty cavity state is given by Eq.~\eqref{eq:cat}, with $\phi_0=0$ and $A=3, 4$ and $5$, indicated by the corresponding initial cavity photon number at the origin $t=0$ of each trajectory. Vertical strokes coincide with APD firings triggering the current sample making. The three insets on the right [inset (i) in the bottom panel plot] depict the normalized transient intensity-field correlation function $h_{\pi/2}(0;\tau)$ (the distributions peak at unity) with different colors either side of the $\tau=0$ axis, and $\kappa \tau_{\rm max}=2$. Inset (ii) in the bottom plot shows $h_{\pi/2}(0;\tau)$ (time window with $\kappa \tau_{\rm max}=4$) calculated from a different realization.  Horizontal lines mark the zero level.}
\label{fig:FIG5}
\end{figure*} 

\subsection{State stability post localization}

We will now look closer at another frequently encountered situation when a downward jump is {\it not} followed by a clear decay to the dim state, as pictured in Fig.~\ref{fig:FIG4}(a). This is often the case when the bright state dominates the steady-state solution of the ME~\eqref{eq:MEJC} over the dim [see the $Q$ function $\mathcal{Q}_{\rm ss}(x+iy)$ in inset (i)]. The QND measurement strategy we described above can still be applied, identifying an equal-weight superposition between the bright and unstable states when the transmission and output flux of the auxiliary cavity plunge to near zero for the first time after a bright-state interval [see Fig.~\ref{fig:FIG4}(b)]. It proves more instructive, however, to once again follow the auxiliary cavity field evolution past that point to detect signs warning that a metastable state is {\it not} about to be established. In fact, the complex of connected unstable states confounding the main peaks, observed in the $Q$ function of inset (ii) in Fig.~\ref{fig:FIG4}(a), can be viewed as a precursor of the intense fluctuations that are to ensue; coherent localization does not occur here between well established amplitudes.

 The first indication to be inferred from the auxiliary cavity that a metastable state has not been established in the main one, is the large deviation of the scaled transmission from the value $|\tilde{\alpha}_1|^2/4$ [Fig.~\ref{fig:FIG4}(b)]. This was not the case in the corresponding trajectory of Fig.~\ref{fig:FIG2}(b) where, in addition, more continuous oscillations of a well defined phase were recorded. The temporal coherence manifested past $t_{\rm dip}$ serves then as a second criterion to distinguish intense fluctuations from an actual jump to a metastable state. Interestingly, we also note that the $y$-quadrature amplitude of the main cavity field does not return to its bright-state value during the period of intense fluctuations, lasting for about four average main cavity lifetimes, despite the fact that the conditioned photon number occasionally does so [compare the trajectories of Figs.~\ref{fig:FIG4}(a) and~\ref{fig:FIG4}(c)].
 
Carrying on with our strategy which dictates cancelling the JC evolution past $t_{\rm dip}$, in Fig.~\ref{fig:FIG4}(d) we show three sample realizations of the integrated charge collected under mode-matched balanced homodyne detection, solving Eq.~\eqref{eq:SDE2}. They are obtained upon setting the LO phase to $\theta=\pi/2$ relative to the axis joining $\alpha_1-\alpha_2$ and its opposite in phase space (once again in the ideal-case scenario where the conditioned amplitudes remain the same across different realization and $A$ is deterministic). All trajectories begin from zero but end up at different points with different signs when all photons have left the empty main cavity, trapped in different wells of the potential~\eqref{eq:Vhom}. The characteristic timescale for such phase localization to be established, working against the phase diffusion in the conditioned wavefunction~\cite{Carmichael1999}, is $\eta_m=1-1/(2A^2)$~\cite{Carmichael1999, CatWPD}. The long-time histogram evinces the familiar interference fringes [Fig.~\ref{fig:FIG4}(e)]. Direct comparison with Fig.~\ref{fig:FIG3}(c) reveals a larger fringe spacing, whence a smaller value of $A$, originating from a bistable switch in which the bright and unstable state amplitudes had a relatively smaller separation in the phase-space {\it quasi}probability distribution.

\section{Wave/particle correlations from macroscopic jumps}
\label{sec:wp}

So far we have remained focused on the integrated charge produced as a measured signal. The last part of our discussion briefly touches on an alternative perspective, one that focuses on single realizations and examines the evolution of the conditioned intracavity photon number during the free decay of the main cavity mode. This quantity, equal to $\langle \psi_{\rm REC}(\eta)|a^{\dagger}a|\psi_{\rm REC}(\eta)\rangle$, reflects the contextual evolution of the cavity state subject to Eq.~\eqref{eq:SSE} (see also~\cite{Carmichael1999}). When the LO phase is set to $\theta=0$ (to measure ``position''), the trajectories evince a monotonic decay $\propto e^{-2\kappa t}$, translated to the straight line with constant negative slope against $\eta$ we observe in Fig.~\ref{fig:FIG4}(f). The dynamical evolution drastically changes when we instead measure ``momentum'' ($\theta=\pi/2$): strong deviations from the exponential decay profile are noted close to $\eta_m$ when the photon number is expected to fall below unity. Moreover, the inset of Fig.~\ref{fig:FIG4}(f) shows a modulation of the cavity excitation, characteristic of the increasing depth in the potential wells of $V(Q_{\pi/2},\eta)$ for $\eta \gtrsim \eta_m$ we saw in the inset of Fig.~\ref{fig:FIG4}(d). The oscillations of $Q_{\pi/2}$ about a given potential well are thus translated to a modulation of the conditional probability of photon emission from the damped empty cavity mode, in spite of the significant depletion of the resonator energy (average intracavity photon number) at these times. 

Let us look further into the dynamical evolution of single quantum trajectories unraveling the ME of an empty cavity mode initiated in the conditional state of a macroscopic bistable switch. Prompted by the contextual departure of sample realizations from a monotonic exponential decay, to initiate Stage 2 we now substitute the BHD of Fig.~\ref{fig:FIG1} with the wave/particle correlator originally introduced in~\cite{Carmichael2000} and pictured in Fig.~\ref{fig:FIG5}(a). In contrast to the complementary balanced detection schemes used in Sec.~\ref{sec:compmethods}, here only a fraction of the cavity output light reaches the homodyne detector, while the rest goes to a photon counter which triggers the photocurrent sampling. The device cross-correlates a photoelectron count with the photocurrent produced by a balanced homodyne detector, in what makes a natural extension of the Hanury-Brown and Twiss technique~\cite{Brown1956, BrownHTwissI, BrownHTwissII}. The accomplished {\it conditional} homodyne detection~\cite{YurkeStoler1987,Schleich1991,Carmichael2000,Wiseman2002, CarmichaelFosterChapter, Carmichael2004,MarquinaCruz2008} proves very sensitive to fluctuations of nonlinear quantum radiation signals~\cite{Foster2001}, while exhibiting only a signal-to-noise dependence on the detection efficiency~\cite{Carmichael2000, CarmichaelFosterChapter}.

The LO phase is again set perpendicular to the line joining $\alpha_1-\alpha_2$ and its opposite in phase space, and $A=|\alpha_1-\alpha_2|/2$. The substitute is used to operationally determine the intensity-field correlations in the transient, modelled with explicit reference to the initial coherent superposition state through the function:
\begin{equation}
h_{\pi/2}(t=0;\tau) \equiv \frac{\sum_{j=1}^{N_s}\langle \mathcal{A}_{\pi/2}(t_j + \tau)\rangle_{\rm REC}}{\sum_{j=1}^{N_s}\langle \mathcal{A}_{\pi/2}(t_j)\rangle_{\rm REC}}.
\end{equation} 
Sampling of the conditioned field quadrature amplitude is triggered by counts out of the APD at the ``start'' times $t_j$. Averaging over $N_s$ samples, corresponding to the set $\{ t_j\}$ and collected in the course of a given realization, shapes a complementary picture of the quadrature amplitude fluctuations of the input field~\cite{Reiner2001, CarmichaelFosterChapter}. Since the cavity mode decays undriven to the vacuum, $N_s$ is a random variable whose distribution explicitly features the initial cavity excitation ($\sim A^2$); histograms over an ensemble of realizations point to a Poisson distribution with mean $0.5A^2$, with contextual deviations noted for different settings of $\theta$~\cite{CatWPD}. The associated photoelectron ``clicks'' are indicated by the vertical strokes in the photon number decay trajectories of Fig.~\ref{fig:FIG5}(b), while the generated time-symmetric correlation functions are pictured as insets at the top right of conditional photon number averages. 

For an input cat state with $A=3$, linked to lower-amplitude bistability, only a bunched pair of ``start clicks'' are registered. The two ``start'' times are centred about the characteristic time of phase localization, $\kappa t_m=(1/2)\ln(2A^2) \approx 1.45$, while the couple of collected samples of the conditioned quadrature amplitude evolution show large relative fluctuations of different sign. The large fluctuations resolve a de-stabilization of phase localization at a timescale set by the two closely spaced emissions. Next, the record obtained with $A=4$ contains eight ``start clicks'', the mean $0.5A^2=8$ of a Poisson distribution. The exponential profile of the decay is nevertheless disturbed by the last (ninth) ``click'' well separated from the rest. More importantly, the derived intensity-field correlation function shows an obvious bias towards evolving fluctuations of the opposite sign relative to their value at zero delay. Finally, the number of samples collected for $A=5$ (totaling $N_s=20$) exceed two standard deviations above the mean of a Poisson distribution, $0.5A^2 + 2\sqrt{0.5 A^2}$. Moreover, the plateau established after the last ``start click'' and fading away after $\kappa t_m \approx 1.96$, resolves once again the occurrence of phase localization, as does the temporal coherence evinced in the oscillatory decaying quadrature amplitude fluctuations---now visibly balanced and reduced in magnitude [inset (i)]. Large and opposite sign deviations from the initial value may still occur [inset (ii)], depending on the extent of charge and field diffusion resolved by the last ``start click'' before the trajectory is trapped into a potential well.

\section{Concluding remarks}

In conclusion, we have performed an operational tomography of quantum jumps between macroscopic and metastable states of light, evaluating the statistical properties of an ensemble of occurrences along bistable realizations, while the corresponding master equation---solved by the ensemble average over those trajectories---evidently exhibits no bistability. We have relied on the spontaneous preparation of a coherent-state superposition following a downward switching event. In the bad cavity limit of an auxiliary resonator, the null measurement record~\cite{Carmichael1993QTI, Carmichael1993QTII, CarmichaelBook2} underlying coherent-state localization~\cite{Carmichael2013Ch4} in the main cavity is the key to produce an unequivocal dip in the transmission of the auxiliary cavity as a coherent cancellation. Therefore, the auxiliary cavity field is the superposition of {\it two coherent drives} in the duration of localization $\Delta t_{\rm loc} \ll \kappa^{-1}$, which itself nonlinearly depends on the balance between drive and photon loss in a paradigmatic out-of-equilibrium quantum phase transition. We point out that the scheme discussed here only constitutes an illustrative example of a potentially general class of configurations taking into account the main field coherence during a macroscopic switch; key to them all will be the time evolving macroscopic superposition of coherent states within a set time window --- the correlation time $\Delta t_{\rm loc}$ of a bound state pair --- as a drive to an auxiliary resonator  or, equally, an `atomic' transition. 

For individual realizations, the mechanism of coherent localization underlying a quantum jump and defining its minimum duration can be compared against the occurrence of jumps in open quantum systems that interact with quantum reservoirs which exhibit non-Markovian dynamics~\cite{Piilo2008, Paavola2009, Groblacher2015, Ghosh2024}. In the ensemble-average sense, the shift from single-atom bistability in its strong-coupling limit ($g/\kappa \gg 1$) to the linearized fluctuations of the many-atom weak coupling limit ($g/\kappa < 1$) of the open JC model could be assessed against the coherence-to-decoherence transition resolved by quantum transport in extended systems~\cite{Haycock2000, Bandyopadhyay2020, Bandyopadhyay2021, Tong2025}. 

The strategy we developed in this report may be employed to explore the absence of detailed balance in the development of quantum amplitude bistability, already deduced at the level of stochastic Maxwell--Bloch equations~\cite{Denisov2002}. At the quantum level, there is no coherent localization in upward jumps; intensely bunched photon emissions condition single-state cavity fields as shown in the {\it quasi}probability distributions of~\citep{JumpsBist}; each photoelectron recorded induces a higher conditional probability for the emission of the next photon until a maximum is reached before the bright state is established. While the path taken to attain a bright state can be tracked in detail by the heterodyne current records, no interference fringes are expected to be seen in the homodyne current histograms. Apart from the interference effect imprinted on the integrated charge, the contextual wave/particle correlations read from individual macroscopic downward jumps uncover operational signatures of quantum coherence through the competition of phase diffusion and localization. Overall, the method suggested is able to tell apart the asymmetric temporal fluctuations that establish metastable states in the breakdown of photon blockade. It is thus sensitive to the exact sequence of individual photoelectron ``clicks'' that make a scattering record of quantum amplitude bistability, and illustrates the correlation of this sequence with the variation of quadrature phase amplitudes.

The data and codes underlying this work are openly available on figshare at this \href{https://doi.org/10.17045/sthlmuni.31366063}{DOI}.

\bibliography{bibliography_Exp_P}

\end{document}